\documentclass[fleqn,usenatbib]{mnras}
\usepackage{newtxtext,newtxmath}
\usepackage{float}
\usepackage{color}
\usepackage{xcolor}
\usepackage{booktabs}
\usepackage{multirow}
\usepackage{natbib}
\usepackage{comment}
\usepackage[normalem]{ulem}
\usepackage[T1]{fontenc}
\usepackage{ae,aecompl}
\usepackage{gensymb}

\usepackage{graphicx}	
\usepackage{amsmath}	
\usepackage{amssymb}	
\usepackage[export]{adjustbox}[2011/08/13]

\newcommand{\Ms}{M$_\odot$}
\newcommand{\bhb}{{\rm BHB}}

\newcommand{\Zs}{Z$_\odot$}
\newcommand{\nbody}{\texttt{NBODY7}}

\newcommand{\argdf}{\texttt{ARGdf }}
\newcommand{\archain}{\texttt{ARCHAIN }}

\title[Stellar Black Hole Binary Mergers in Open Clusters]{Stellar Black Hole Binary Mergers in Open Clusters}

\author[S. Rastello et al.]{
S. Rastello$^{1}$\thanks{E-mail: sara.rastello@uniroma1.it},
P. Amaro-Seoane$^{2}$,
M. Arca-Sedda$^{1,\,3}$,
R. Capuzzo-Dolcetta$^{1}$,\newauthor
G. Fragione$^{4}$,
I. Tosta e Melo$^{1}$
\\
$^{1}$Dep. of Physics, Sapienza, Universit\'a di Roma, P.le A. Moro 5, 00185, Roma, Italy\\
$^{2}$Institute of Space Sciences (ICE, CSIC) \& Institut d'Estudis Espacials de Catalunya (IEEC)\\
     at Campus UAB, Carrer de Can Magrans s/n 08193 Barcelona, Spain\\
Institute of Applied Mathematics, Academy of Mathematics and Systems Science, CAS, Beijing 100190, China\\
Kavli Institute for Astronomy and Astrophysics, Beijing 100871, China\\
Zentrum f{\"u}r Astronomie und Astrophysik, TU Berlin, Hardenbergstra{\ss}e 36, 10623 Berlin, Germany\\
$^{3}$Astronomisches Rechen-Institut,  M\"onchhofstra\ss{}e 12-14, 69120 Heidelberg, Germany\\
$^{4}$Hebrew University of Jerusalem, Racah Institute of Physics, Givat Ram, 9190401, Jerusalem, Israel
}

\date{Accepted XXX. Received YYY; in original form ZZZ}

\pubyear{2018}
\begin{document}
\label{firstpage}
\pagerange{\pageref{firstpage}--\pageref{lastpage}}
\maketitle

\begin{abstract}
%
%
%
%
%
%
In this paper we study the evolution of a primordial black hole binary (BHB) in a sample of over 
1500 direct-summation $N-$body simulations of small-and intermediate-size isolated star clusters as proxies of galactic open clusters. The BHBs have masses in the range of the 
first LIGO/Virgo detections. Some of our models show a significant hardening of the BHB in a relatively short time. Some of them merge within the cluster, while ejected binaries, typically, have exceedingly long merger timescales. The perturbation of stars around BHB systems is key to induce their coalescence. The BHBs which merge in the cluster could be detected with a delay of a few years between space detectors, as future LISA, and ground-based 
ones, due to their relatively high eccentricity. Under our assumptions, 
we estimate a BHB merger rate of $R_{\rm mrg} \sim 2$ yr$^{-1}$ Gpc$^{-3}$.  
We see that in many cases the BHB triggers tidal disruption events which, however, 
are not linked to the GW emission.
Open cluster-like systems are, hence, a promising environment for GWs from BHBs
and tidal disruptions. 
\end{abstract}




\section{Introduction}

The Laser Interferometer Gravitational-Wave Observatory (LIGO) and Virgo have
detected six sources of GWs as of writing this article
\citep{abbott16a,abbott16b,abbott17,abbott17a,abbott17b,abbott17c}. All of them
but one, GW170817, last reference, correspond to a system of BHB. Typically,
the masses are larger than the nominal one derived from stellar evolution of
$10\,M_{\odot}$, as predicted by \cite{Amaro-SeoaneChen2016}, i.e.
``hyperstellar-mass black holes'', as the authors coined,
 and previously discussed by e.g. also
 \citealt{heger2003,mapelli08,zampieri09,mapelli10,belczy2010,fryer12,mapelli13,ziosi2014,speetal15}.\\

There are two different channels to form a BHB, namely either (i) in the field
in isolation and via stellar evolution of a binary of two extended stars
 \citep[see e.g.][]{tutukov,bethe, belczynski2002,belckzynski2010,postnov14,loeb16,tutukov17,postnov17,giacobbo18}.

(ii) or via dynamical interactions in a dense stellar system (see the review of
\citealt{BenacquistaDowning2013} and e.g. also
\citealt{sambaran10,downing2010,mapelli13,ziosi2014,rodriguez15,rodriguez16,mapelli16,askar17,antonini16,megan17,frak18}).

In this article we address the evolution of a BHB in an open cluster with a
suit of 1500 direct-summation $N-$body simulations. We model the evolution of
the BHB with properties similar to what can be expected to be detected by
LIGO/Virgo \citep{Amaro-SeoaneChen2016}, using different representations of
small- and intermediate-mass isolated open star clusters.

Our cluster models are considered as a proxy of the Galactic population of open
clusters, which are characterized by central densities of a few M$_\odot\,{\rm
pc}^{-3}$, i.e. much lower than the typical densities of globular clusters, and
contain a number of stars from two to three orders of magnitude less than a
globular cluster.

We investigate the evolution of stellar BHBs in low-mass and low-density open
clusters models by means of high precision direct-summation $N-$body
simulations.  In an open cluster the impulsive effect produced by the large
fluctuations over the mean field, whose amplitude is of order $\sqrt{N}/N$, can
significantly affect the BHB evolution.  Assuming an initial number of star
$N_{\rm o}=1000$ for this type of open clusters and $N_{\rm g}=10^6$ for a
typical globular cluster, we can calculate the expected fluctuations over the
mean field amplitude as $f=\sqrt[]{N_{\rm g}/N_{\rm o}}=\sqrt[]{1000}\simeq
32$, thus implying a larger such effect in open clusters. This enhanced effect
of stochastic fluctuations (physically given by the rare but close approaches
among cluster stars) reflects in the ratio of the 2-body relaxation time scales
which, given the cluster sizes as $R_{\rm o}$ and $R_{\rm g}$, writes as
\citep{spi87}

\begin{equation}
\frac{t_{\rm rlx,\,o}}{t_{\rm rlx,\,g}}=\frac{1}{f}\frac{\log(0.11\,N_{\rm o})}{\log(0.11\,N_{\rm g})}\left(\frac{R_{\rm o}}{R_{\rm g}}\right)^{3/2}.
\end{equation}

Assuming $R_{\rm o}/R_{\rm g} = 1/5 $, the above equation yields to $t_{\rm
rlx,\,o}/t_{\rm rlx,\,g}\simeq 0.02$, meaning that the smaller system evolves
50 times faster.  Of course, this enhanced effect of individual strong
encounters is partly compensated by their smaller time rate.

In this paper we address the evolution of a BHB which, as a result of dynamical
friction orbital decay \citep[see e.g.][]{bt},  we assume to be located at the
centre of an open cluster-like system. The masses of the black holes are set to
$30\,M_{\odot}$ each, following the first LIGO/Virgo detection, the GW150914
source \citep{abbott16a}, and \cite{Amaro-SeoaneChen2016}. Despite the possible
ejection due to the supernova natal kick, there is margin for such kind of
remnant to be retained in an open cluster. Indeed, compact remnants such as
neutron stars and black holes formed in massive binaries are much easier
retained in clusters because the kick momentum is shared with a massive
companion, which leads to a much lower velocity for the post-supernova binary
\citep{podsi04,podslia2005}. In the case of neutron stars, \cite{podsi04}
showed that for periods below 100 days, the supernova explosion leads to a very
little or no natal kick at all (their Fig.2, the dichotomous kick scenario).
Open clusters have binaries mostly with periods of 100 days and below (see
\citealt{Mathieu2008}, based on the data of \citealt{DuquennoyMayor1991}).
These results can be extrapolated to black holes because they receive a similar
kick to neutron stars (see \citealt{RepettoEt2012} and also the explanation of
\citealt{Janka2013} of this phenomenon). In any case, black holes with masses
greater than $10\,M_{\odot}$ at solar metallicity form via direct collapse and
do not undergo supernova explosion, and hence do not receive a natal kick
\citep{PernaEtAl2018}.  Also, while the solar metallicity in principle could
not lead to the formation of black holes more massive than $25\,M_{\odot}$
\cite{spema17}, we note that the resonant interaction of two binary systems can
lead to a collisional merger which leads to formation of this kind of black
hole \citep{goswami2004,fregeau2004} at the centre of a stellar system, where
they naturally segregate due to dynamical friction.

Moreover, we note that another possibility is that these stellar-mass black
holes could have got their large masses due to repeated relativistic mergers of
lighter progenitors. However, the relativistic recoil velocity is around
$200-450\,{\rm km/s}$ for progenitors with mass ratio $\sim [0.2,\,1]$
respectively, so that this possibility is unlikely \citep[see e.g.][their Fig.
1, lower panel]{Amaro-SeoaneChen2016}, because they would escape the cluster,
unless the initial distribution of spins is peaked around zero and the black
holes have the same mass (as in the work of \cite{RodriguezEtAl2018} in the
context of globular clusters). In this case, second generation mergers are
possible and, hence, one can form a more massive black hole via successive
mergers of lighter progenitors.

The relatively low number of stars of open clusters gives the possibility to
integrate over at least a few relaxation times in a relatively short
computational time, so that, contrarily to the cases of globular clusters or
galactic nuclei, it is possible to fully integrate these systems without the
need to rescale the results.

In this article we present a series of 1500 dedicated direct-summation $N-$body
simulations of open clusters with BHBs.  The paper is organized as follows: in
Sect. 2 we describe the numerical methods used and our set of models; in Sect.3
we present and discuss the results of the BHB dynamics; in Sect. 4 we discuss
the implication of our BHBs as sources of gravitational waves; in Sect. 5 we
present the results on tidal disruption events, in Sect. 6 we draw overall
conclusions.

\section{Method and Models}
\label{met_mod}

To study the BHB evolution inside its parent open cluster (henceforth OC) we
used \texttt{NBODY7} \citep{aarsethnb7}, a direct N-body code that integrates
in a reliable way the motion of stars in stellar systems, and implements a
careful treatment to deal with strong gravitational encounters, taking also
into account stellar evolution.  We performed several simulations at varying
both OC and BHB main properties, taking advantage of the two high-performance
workstations hosted at Sapienza, University of Roma, and the Kepler cluster,
hosted at the Heidelberg University.

\begin{table*}
\centering
\caption{
Main parameters characterizing our models. The first two columns refers to the
cluster total number of stars, $N_{\rm cl}$, and its mass $M_{\rm cl}$. The second
two-column group refers to the BHB parameters: semi-major axis, \textit{a}, and
initial eccentricity, \textit{e}. The last column gives the model
identification name. Each model is comprised of 150 different OC realizations.
}
\label{ictab}
\begin{tabular}{@{}ccccccc@{}}
\toprule
\multicolumn{2}{c}{\textbf{Cluster}}                           &  & \multicolumn{2}{c}{\textbf{BHB}} & \textbf{} & \textbf{$N$-body set} \\ \midrule
$N_{\rm cl}$                     & $M_{\rm cl}$ (\Ms)                                &  &  \textit{a} (pc)                     & \textit{e}   &           & Model               \\ \midrule
\multirow{2}{*}{512}  & \multirow{2}{*}{$ 3.2 \times 10^{2}$} &  & \multirow{2}{*}{0.01}    & 0.0   &           & A00                 \\
                      &                                        &  &                          & 0.5   &           & A05                 \\ \midrule
\multirow{2}{*}{1024} & \multirow{2}{*}{$ 7.1 \times 10^{2}$} &  & \multirow{2}{*}{0.01}    & 0.0   &           & B00                 \\
                      &                                        &  &                          & 0.5   &           & B05                 \\ \midrule
\multirow{2}{*}{2048} & \multirow{2}{*}{$ 1.4\times 10^{3}$} &  & \multirow{2}{*}{0.01}    & 0.0   &           & C00                 \\
                      &                                        &  &                          & 0.5   &           & C05                 \\ \midrule
\multirow{2}{*}{4096} & \multirow{2}{*}{$ 2.7 \times 10^{3}$} &  & \multirow{2}{*}{0.01}    & 0.0   &           & D00                 \\
                      &                                        &  &                          & 0.5   &           & D05                 \\ \bottomrule
\end{tabular}
\end{table*}

Table \ref{ictab} summarizes the main properties of our $N$-Body simulation
models.  We created four simulation groups representing OC models at varying
initial number of particles, namely $512\leq N \leq 4096$.  Assuming a
\citet{kroupa01} initial mass function ($0.01$ \Ms $\leq$ M $\leq$
$100$ \Ms), our OC model masses range between $ 300$ M$_{\odot}$ and $3000$
\Ms.  All clusters are modeled according to a Plummer density profile
\citep{Plum} at virial equilibrium with a core radius ($r_c = 1$ pc), and
adopting solar metallicity (\Zs).
We perform all the simulations
including the stellar evolution recipes that are implemented in the \nbody $\,$ code
which come from the standard \texttt{SSE} and \texttt{BSE} tools
\citep{hurley2000, hurley2002}, with updated stellar mass loss and remnant formation prescriptions from \citet{bel2010}.
Further, for simplicity, we do not take into account primordial binaries, which we leave to future work.
 To give statistical significance to the
results we made $150$ different realizations of every model, which are denoted
with names A00, A05, B00, B05, C00, C05, D00 and D05, where the letter refers
to increasing $N$ and the digits to the initial BHB orbital eccentricity.
Additionally, we ran a further sample of $421$ simulations, aiming at
investigating the implications of some of our assumptions on the BHB evolution.
These models are deeply discussed in Sect.  \ref{tde}.

In all our simulations, we assumed that the BHB is initially placed at the
centre of its host OC, and is composed of two equal mass BHs with individual
mass M$_{\rm{BH}}$ = $30$ \Ms.  The initial BHB semi major axis is $0.01$ pc
with two initial eccentricities, $e_{\rm BHB}=0$, $e_{\rm BHB}=0.5$.  The initial
conditions drawn this way are obtained updating the procedure followed in
\cite{ASCD15He}.  The choice of a BHB initially at rest at centre of the
cluster with that not very small separation is not a limitation because the
dynamical friction time scale of $30$ M$_{\odot}$ is short enough to make
likely that both the orbital decay of the BHB occurs rapidly and also that the
probability of a rapid formation of a BHB from two individual massive BHs is
large even on a short time.

The BHB orbital period is, for the given choices of masses and semimajor axis,
P$_{\rm BHB}= 0.012$ Myr.  Note that our BHBs are actually "hard" binaries
\citep{heggie75,hills75,bt}, having binding energy BE$\sim$ 3.8 $10^{45}$ erg,
which is larger than the average kinetic energy of the field stars in each type
of cluster studied in this work.  All models were evolved up to $3$ Gyr, which
is about 3 times the simulated OC internal relaxation time.
The scope of the present work is to give investigate the BHB dynamical evolution, hence we focus on tracking mainly its evolution. We also note that stellar-mass BHs naturally form binary systems in
open clusters over a wide cluster mass range, and can also undergo
triple-/subsystem-driven mergers, as recently shown through explicit direct N-body simulations 
by \citet{kimpson} and \citet{sambaran1}.

\section{Dynamics of the black hole binary}
\subsection{General evolution}
\label{BHB_ev}

The BHB is assumed to be located at the centre of the cluster. Due to
interactions with other stars, the BHB can either undergo one of the following
three outcomes. First, (i) the BHB can shrink and hence become harder, meaning
that the kinetic energy of the BHB is higher than the average in the system
\citep[see e.g.][]{bt}; also (ii) the BHB can gain energy and therefore
increase its semi-major axis and (iii) the BHB can be ionised in a typically
three-body encounter. In Table~\ref{ub} we show the percentages of these
three outcomes in our simulations.

\begin{figure}
\centering
\includegraphics[width=0.5\textwidth]{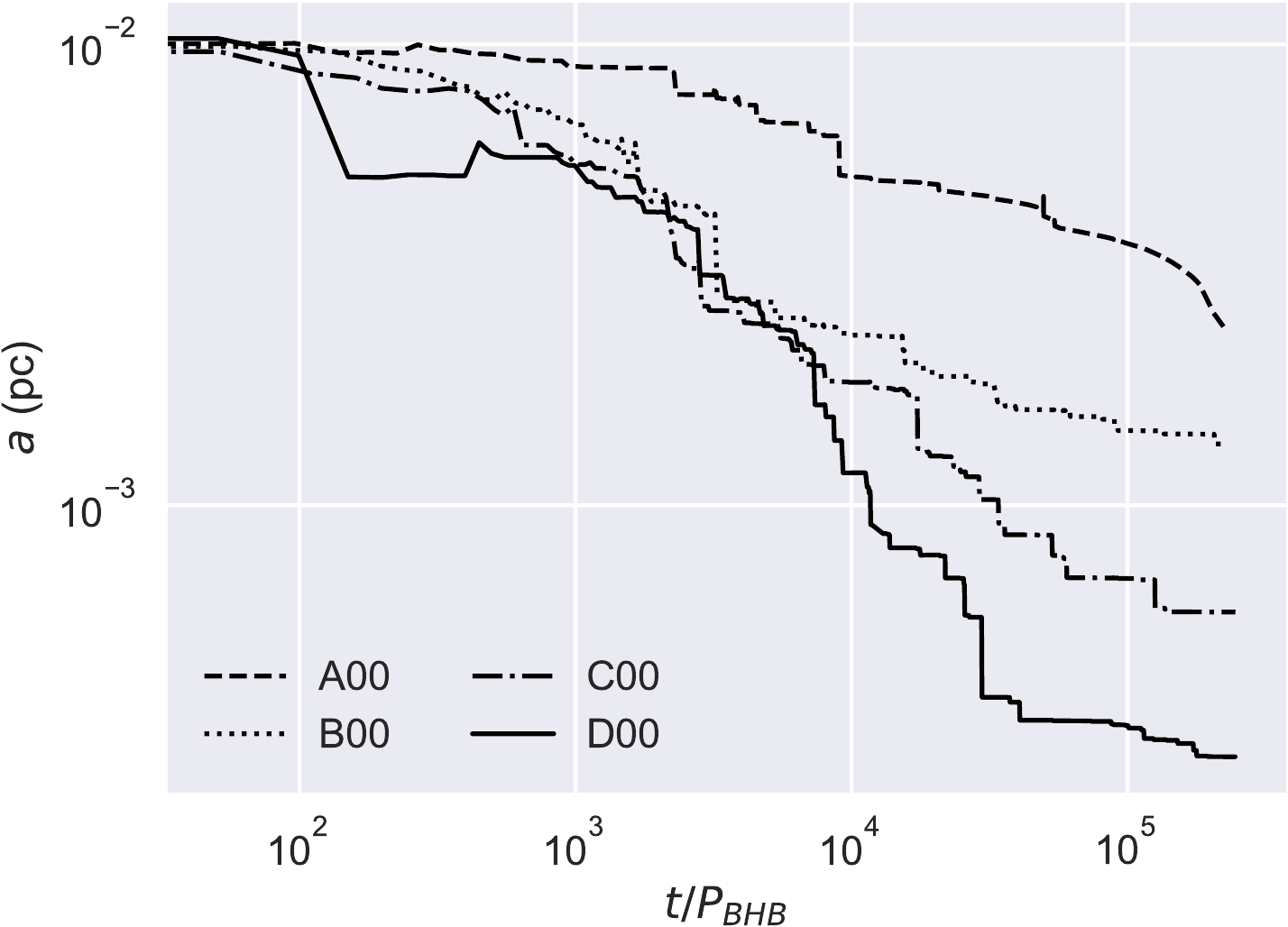}
\includegraphics[width=0.5\textwidth]{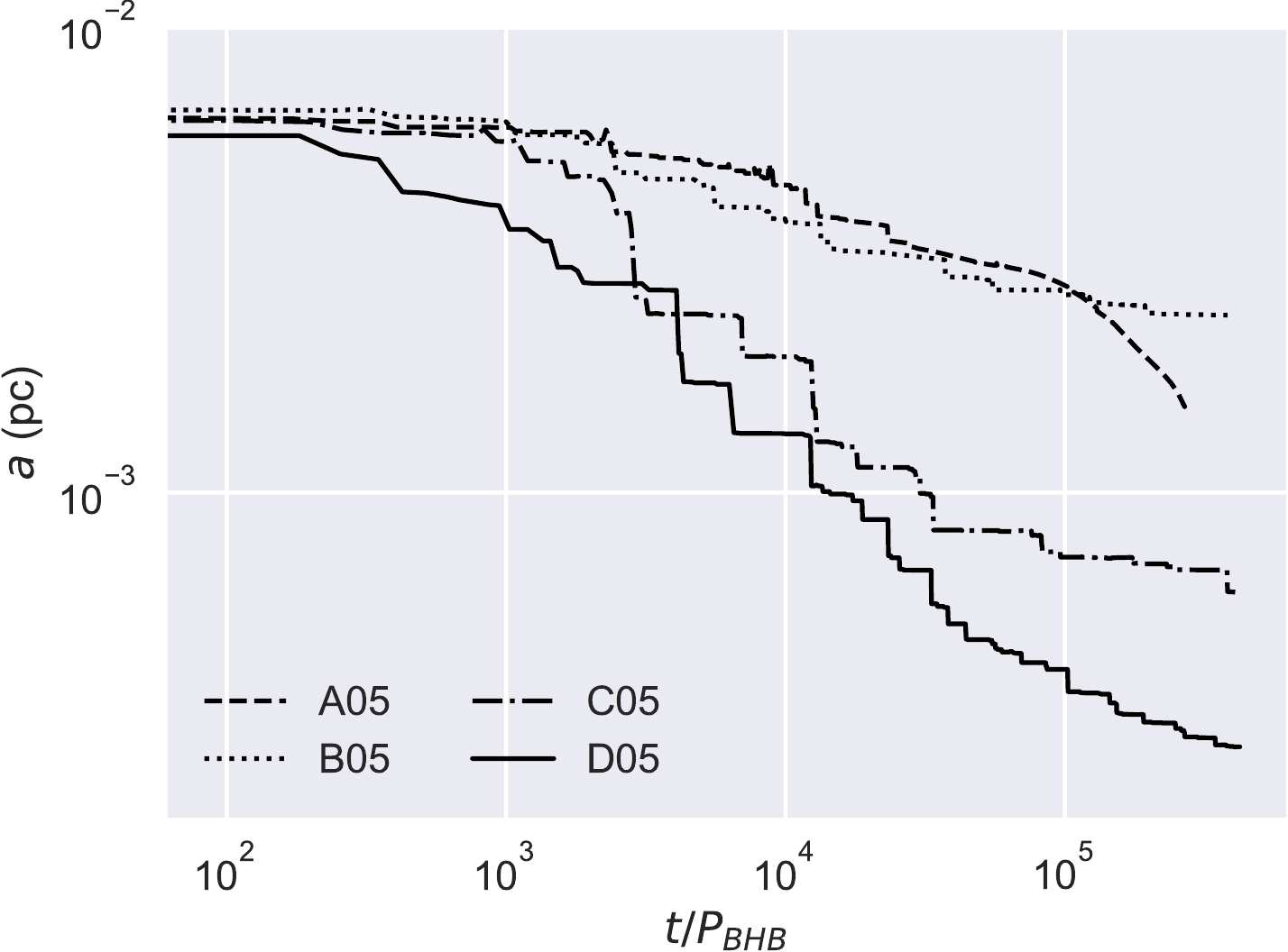}
\caption{
Semi-major axis evolution in four random, representative simulations. The upper panel
shows binaries which initially are circular and lower panel depicts eccentric ones.
We normalise the time to the initial (i.e. at $T=0$) period of the binary.
}
\label{fig:param_sh}
\end{figure}

\begin{table}
\centering
\caption{
Percentage of BHB which undergo one of the three processes described in the
text. They either shrink (column 2), or increase their semi-major axis (column
3) or break up (column 4).
}
\label{ub}
\begin{tabular}{@{}cccc@{}}
\toprule
\textbf{Model} & \textbf{Harder} & \textbf{Wider}& \textbf{Break up} \\
               &            \%        &    \%    &    \%         \\\midrule
A00   & 89.1                  & 7.9                 & 2.9                   \\
A05   & 97.1                  & 2.1                  &   0.7                \\
B00   & 92.5                  & 2.7                  &   4.8                \\
B05   & 94.0                  & 2.0                 &  4.0                  \\
C00   & 93.6                  & 0                   &     6.4            \\
C05   & 96.5                  & 0                   &     3.5             \\
D00   & 94.2                  & 0                  &     5.8              \\
D05   & 97.1                  & 0                   &      2.8            \\ \bottomrule
\end{tabular}
\end{table}

We can see that typically about $90$\% of all binaries shrink their semi-major
axis as they evolve, as one can expect from the so-called \textit{Heggie's law}
\citep{heggie75}.  We note that models in which the binary initially was
eccentric lead to a higher percentage in the ``harder'' outcome.  We display in
Fig.~\ref{fig:param_sh} a few representative examples of these processes.
The decrease is gradual for model A and B while model C and D (which are the
more massive) show a steeper decrease.

There are however cases in which the binary gains energy from gravitational
encounters and increases its semi-major axis, becoming ``wider'' (Table
\ref{ub}, column 2).
If the host cluster is massive enough, the semi-major
axis always decreases (models C and D), contrary to lighter models, in which it
can increase (models A and B).

Because of the initial choice of the BHB semi-major axis, gravitational
encounters with other stars rarely ionise it, although we observe a few events,
typically below $7\,\%$ (circular binaries are easier to ionise). This
ionisation happens between $\sim 5$ Myr and up to $\sim 100$ Myr and it is usually driven
 by the encounter with a massive star ($\gtrsim 10$ \Ms).  In such case,
the massive star generally pair with one of the BHs, while the other BH
is usually ejected from the stellar systems.

\subsubsection{Pericentre evolution}

\begin{figure*}
\begin{minipage}[b]{0.49\linewidth}
\centering
\includegraphics[width=\textwidth]{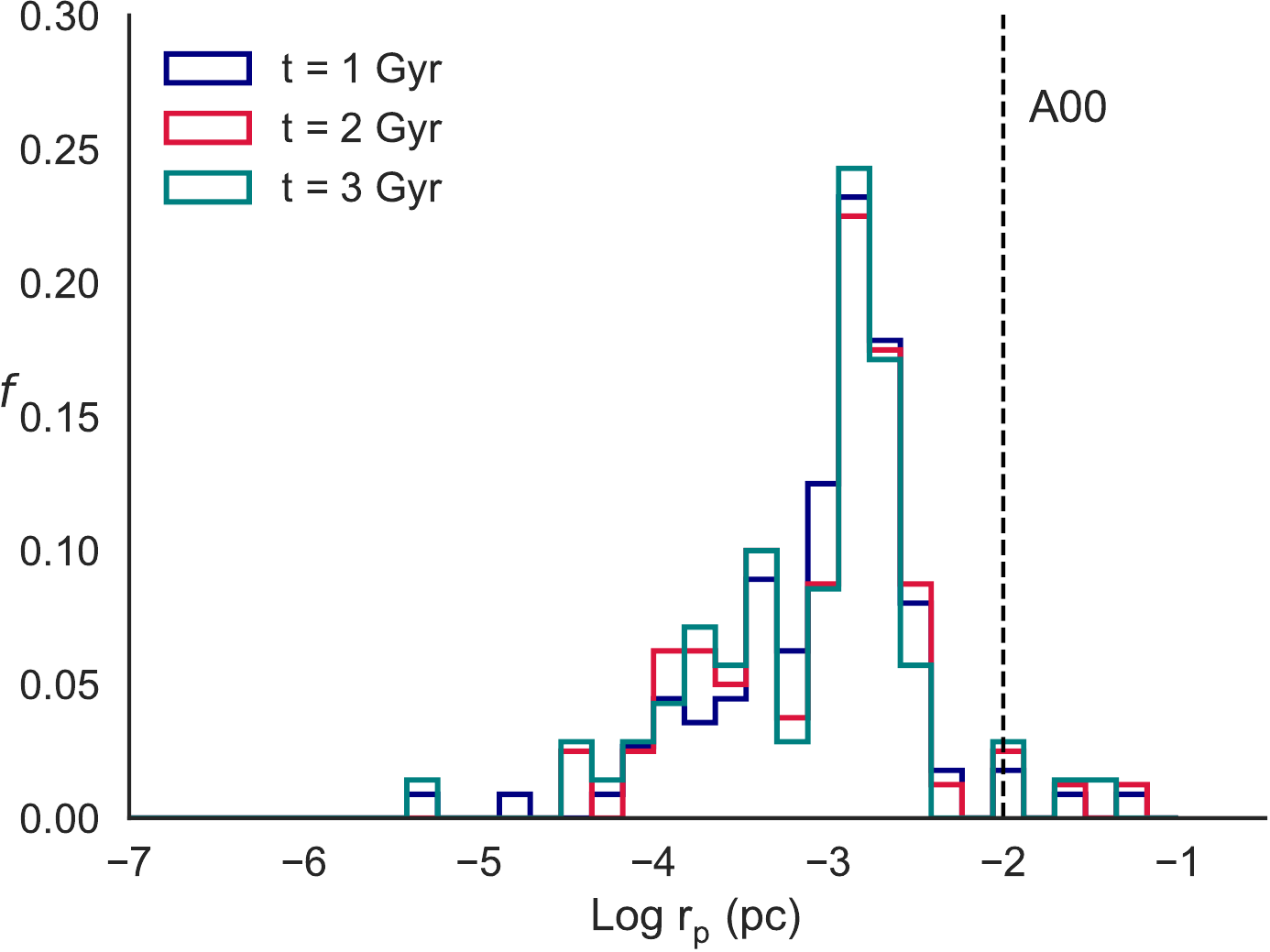}
\end{minipage}
\begin{minipage}[b]{0.49\linewidth}
\centering
\includegraphics[width=\textwidth]{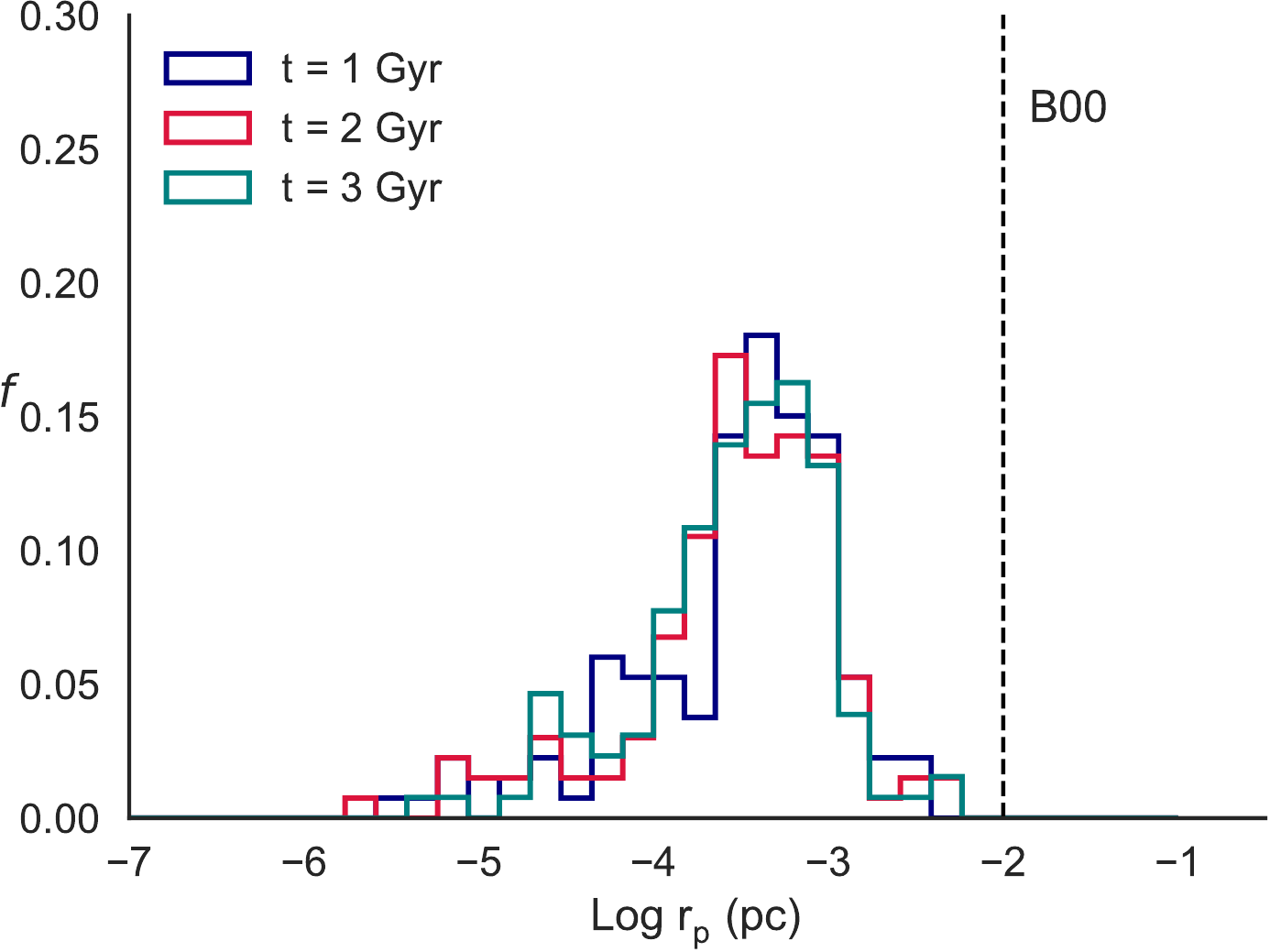}
\end{minipage}
\begin{minipage}[b]{0.49\linewidth}
\centering
\includegraphics[width=\textwidth]{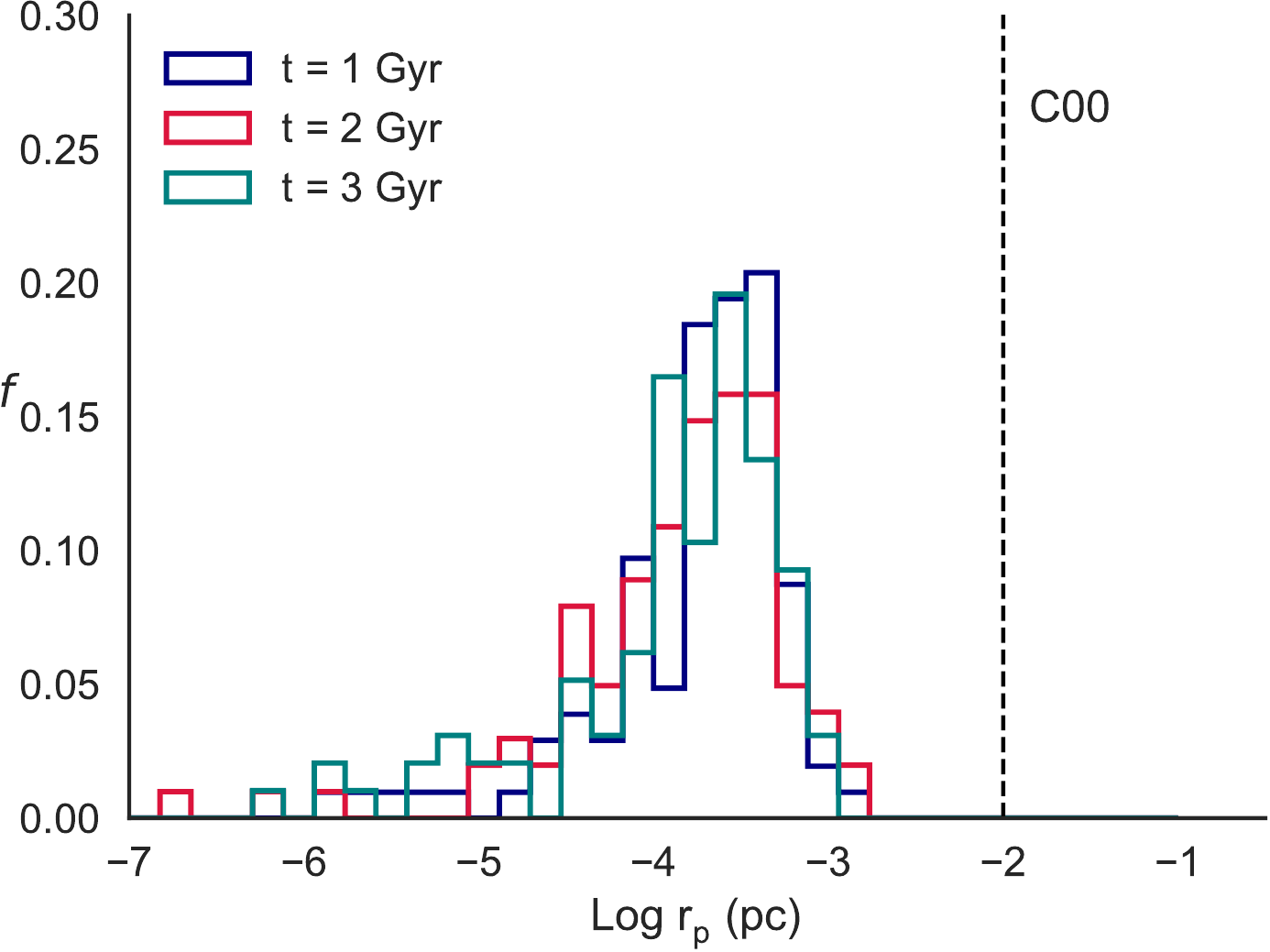}
\end{minipage}
\begin{minipage}[b]{0.49\linewidth}
\centering
\includegraphics[width=\textwidth]{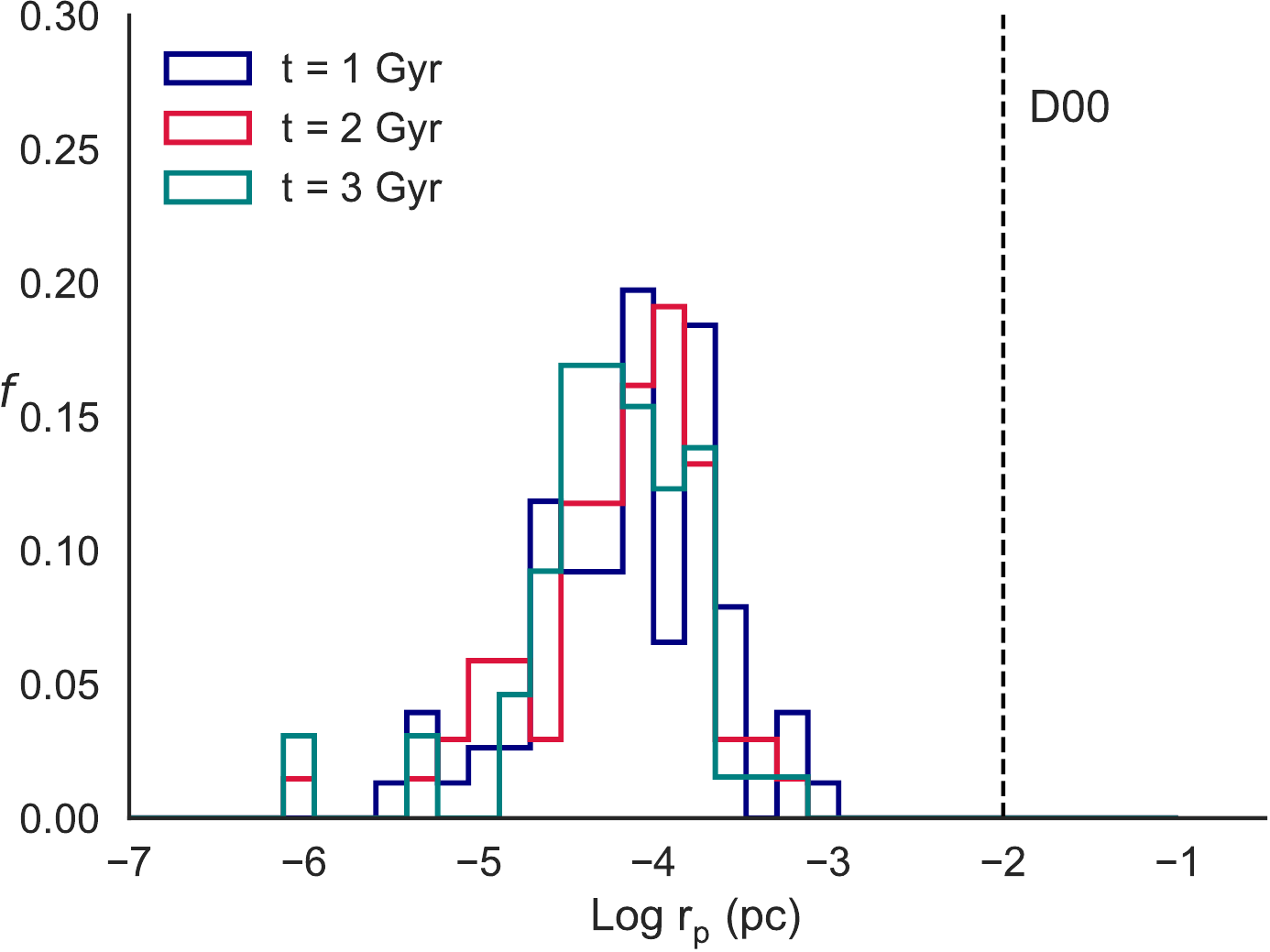}
\end{minipage}
\caption{
BHB pericentre distance distribution for all simulations of models A00, B00,
C00 and D00. The histograms are calculated at three different points in the
evolution of the systems, namely at $1$ Gyr (blue), $2$ Gyr (red) and $3$
Gyr (green). We show with a vertical, black dashed line the initial pericentre
in the model.
}
\label{fig:per_a}
\end{figure*}

\begin{figure*}
\begin{minipage}[b]{0.49\linewidth}
\centering
\includegraphics[width=\textwidth]{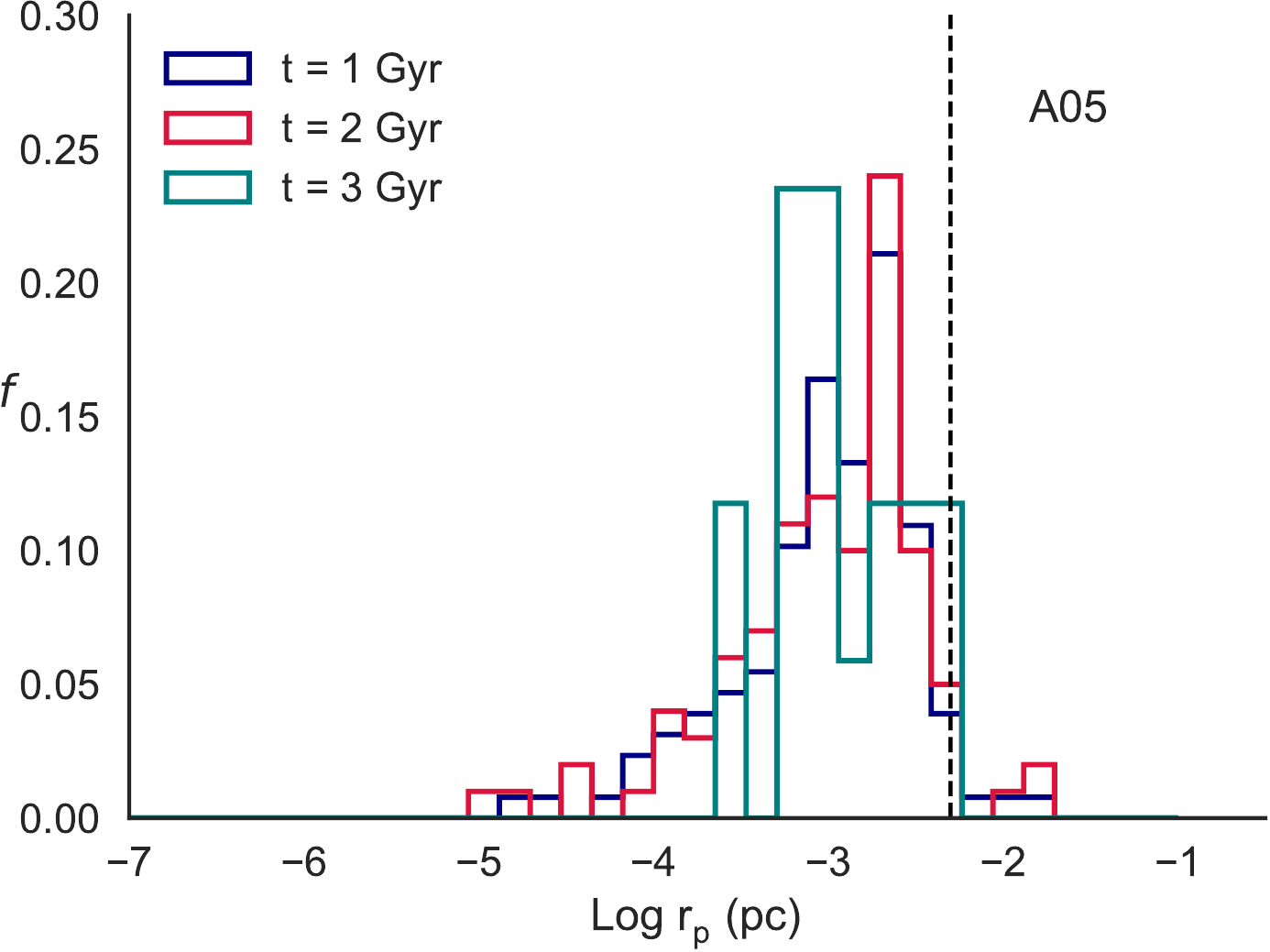}
\end{minipage}
\begin{minipage}[b]{0.49\linewidth}
\centering
\includegraphics[width=\textwidth]{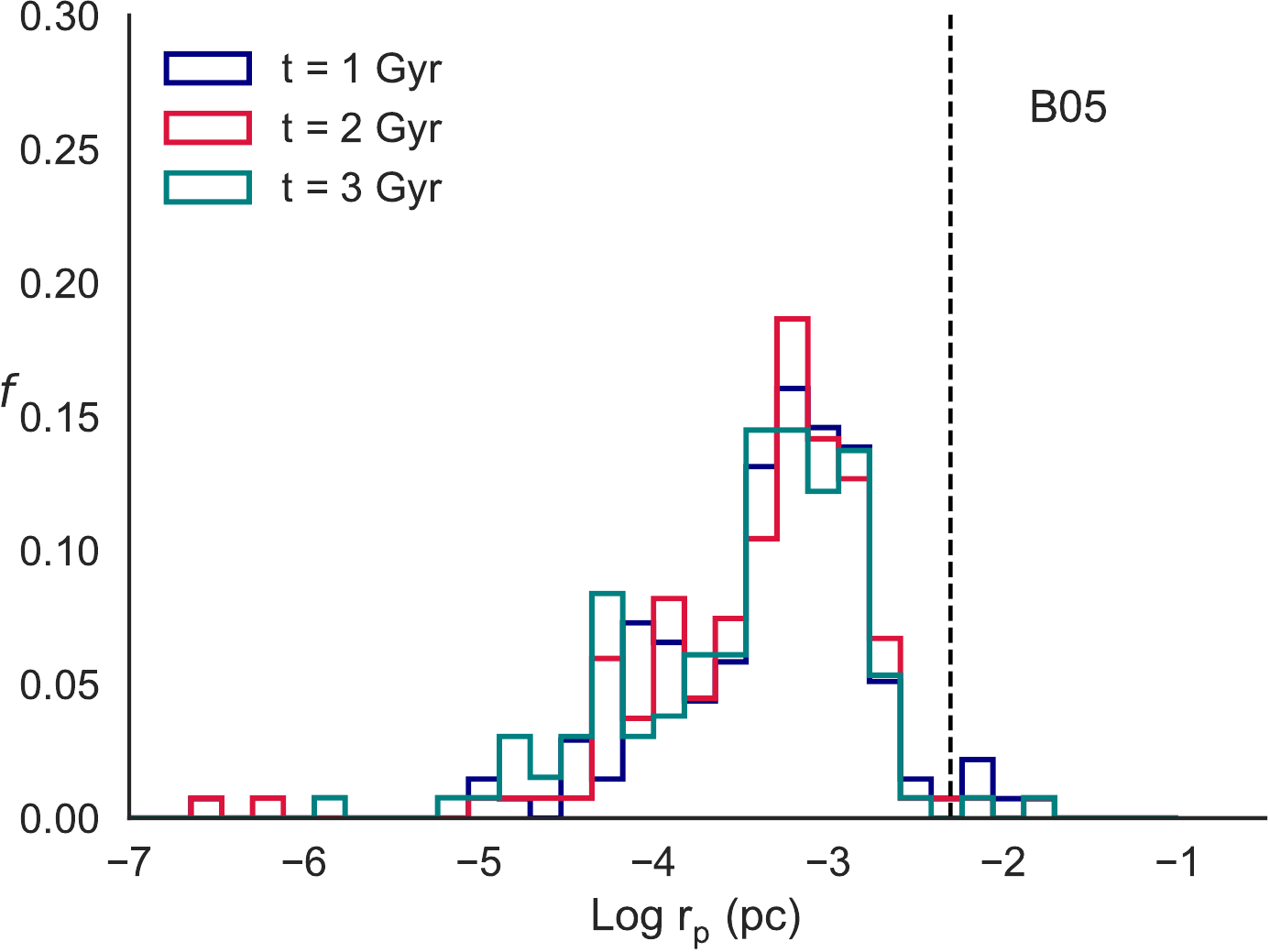}
\end{minipage}
\begin{minipage}[b]{0.49\linewidth}
\centering
\includegraphics[width=\textwidth]{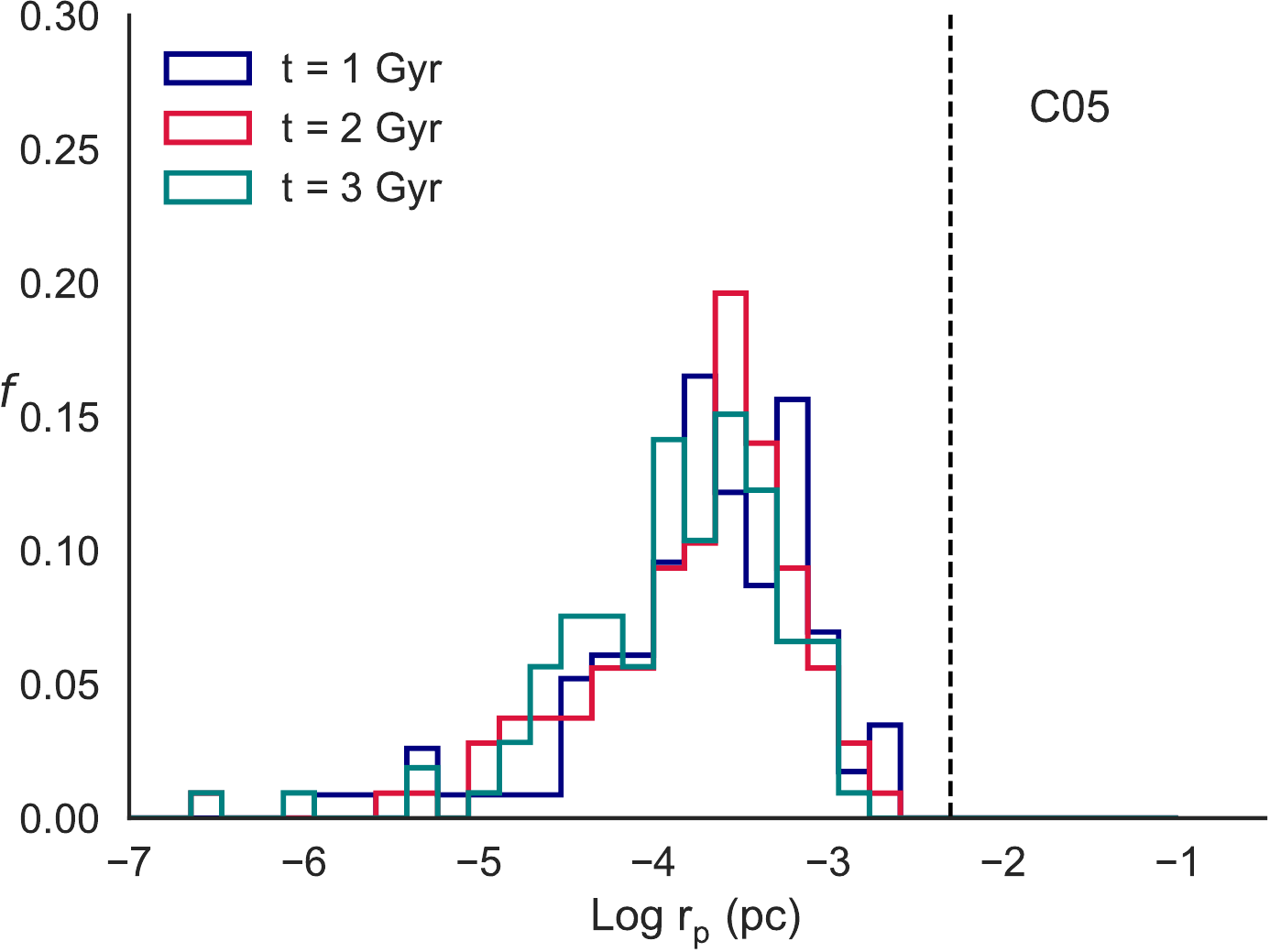}
\end{minipage}
\begin{minipage}[b]{0.49\linewidth}
\centering
\includegraphics[width=\textwidth]{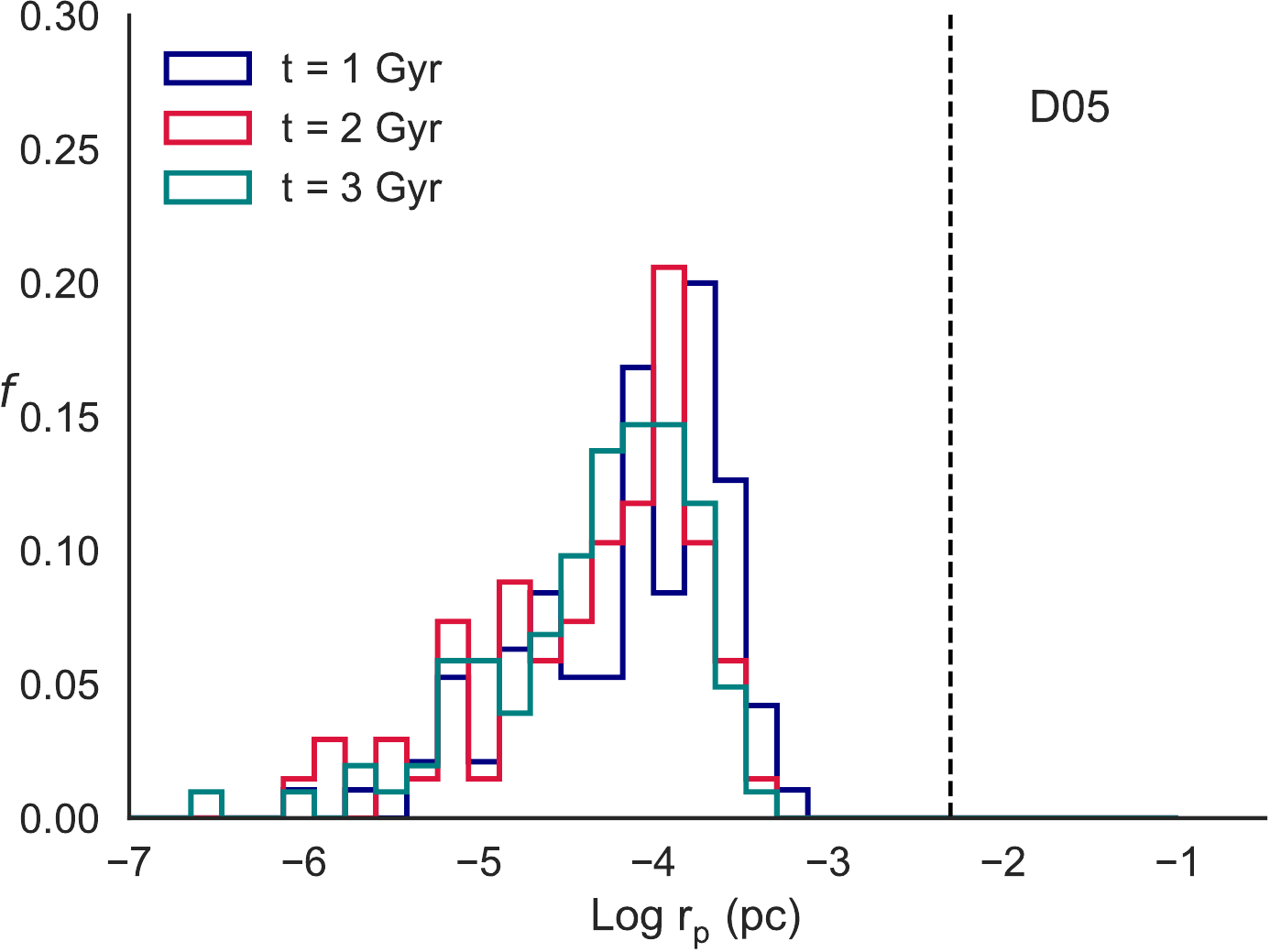}
\end{minipage}
\caption{As in Fig.\ref{fig:per_a}, but for the models A05, B05, C05 and D05.}
\label{fig:per_b}
\end{figure*}

For a BHB to become an efficient source of GWs, the pericentre distance must be
short enough. In this section we analyse the evolution of the pericentres for
our different models. In Table~\ref{peritab} we summarise the results of
Figs.~\ref{fig:per_a}, \ref{fig:per_b}. In the table we show the average
pericentre distance at three different times(1, 2 and 3 Gyr)
in the evolution of the cluster as well as the absolute minimum
pericentre distance we find at each of these times.

\begin{table}
\centering
\caption{
Evolution of the BHB pericentre distance for all the models. The columns from
left to right denote, respectively: the model, the initial BHB pericentre
(r$_{\rm p}^{i}$), the time in which we have calculated the average ($T$), the BHB
pericentre distance averaged over all the simulations of the respective model
($\langle r_{\rm p}\rangle$), and the absolute minimum distance we record
(r$_{\rm p}^{min}$). We note that the Schwarzschild radius of a $30\,M_{\odot}$ is
$1.43\times\,10^{-12}~{\rm pc}$.
}
\label{peritab}
\begin{tabular}{@{}ccccccl@{}}
\hline
\textbf{Model}                & 				\textbf{r$_{\rm p}^{i}$}     & \textbf{T} & \textbf{$\langle r_{\rm p}\rangle$} & \textbf{r$_{\rm p}^{min}$} &  \\
 &      (pc)  &  (Gyr)  &   (pc)  &  (pc)  & \\
\hline
    & & 1   & 			$2.3\times 10^{-3}$ & $5.0\times 10^{-6}$     &  \\
A00  & $1.0\times 10^{-2}$ & 2   &  $2.3\times 10^{-3}$ & $3.2\times 10^{-5}$     &  \\
    & & 3   & 			$2.1\times 10^{-3}$ & $4.9\times 10^{-6}$     &  \\ \midrule
    & & 1   & 			$1.7\times 10^{-3}$ & $1.4\times 10^{-5}$     &  \\
A05 & $5.0\times 10^{-3}$ & 2   & 	$1.9\times 10^{-3}$ & $1.0\times 10^{-5}$     &  \\
    & & 3   & 		 $1.7\times 10^{-3}$ & $2.7\times 10^{-4}$     &  \\ \midrule
    & & 1   & 		 $5.4\times 10^{-4}$ & $3.4\times 10^{-6}$     &  \\
B00 & $1.0\times 10^{-2}$ & 2   & 	$5.7\times 10^{-4}$ & $2.2\times 10^{-6}$     &  \\
    & & 3   & 		 $5.1\times 10^{-4}$ & $4.1\times 10^{-6}$     &  \\ \midrule
    & & 1   & 		 $1.1\times 10^{-3}$ & $2.4\times 10^{-7}$     &  \\
B05 & $5.0\times 10^{-3}$  & 2    & $8.9\times 10^{-4}$ & $2.4\times 10^{-7}$     &  \\
    & & 3   & 		 $7.9\times 10^{-4}$ & $1.5\times 10^{-6}$     &  \\ \midrule
    & & 1   & 		 $2.8\times 10^{-4}$ & $1.7\times 10^{-6}$     &  \\
C00 & $1.0\times 10^{-2}$ & 2     & $2.6\times 10^{-4}$ & $2.2\times 10^{-7}$     &  \\
    & & 3   & 		 $2.5\times 10^{-4}$ & $5.3\times 10^{-7}$     &  \\ \midrule
    & & 1   &		 $3.7\times 10^{-4}$ & $1.5\times 10^{-6}$     &  \\
C05 & $5.0\times 10^{-3}$ & 2   & 	$3.1\times 10^{-4}$ & $2.5\times 10^{-7}$     &  \\
    & & 3   & 		 $2.6\times 10^{-4}$ & $2.5\times 10^{-7}$     &  \\ \midrule
    & & 1   & 		 $1.3\times 10^{-4}$ & $2.7\times 10^{-6}$     &  \\
D00 & $1.0\times 10^{-2}$ & 2   &   $1.0\times 10^{-4}$ & $9.2\times 10^{-7}$     &  \\
    & & 3   &		 $9.1\times 10^{-5}$ & $8.8\times 10^{-7}$     &  \\ \midrule
    & & 1   &        $1.5\times 10^{-4}$ & $1.8\times 10^{-6}$      &  \\
D05 & $5.0\times 10^{-3}$& 2   & 	$1.5\times 10^{-4}$ & $3.6\times 10^{-6}$      &  \\
    & & 3   & 			$1.3\times 10^{-4}$ & $9.8\times 10^{-6}$      &  \\ \midrule
\multicolumn{1}{l}{} & \multicolumn{1}{l}{} & \multicolumn{1}{l}{}                   & \multicolumn{1}{l}{}                    & \multicolumn{1}{l}{}     & \multicolumn{1}{l}{}&
\end{tabular}
\end{table}

For BHB which initially are circular, we can see in the table and in
Fig.~\ref{fig:per_a} that in all models there is a significative shrinkage of
the pericentre distance of, at least, one order of magnitude.
Such shrinkage occurs after only 1 Gyr. For the most
massive clusters, i.e. model D00, about $20\%$ of all binaries achieve a
pericentre distance which is of about two orders of magnitude smaller than the
initial value. We note, however, that a very few binaries shrink to extremely
small values, reaching pericentre distances of down to $10^{-7}\,pc$.
Eccentric binaries also shrink and achieve smaller pericentre values, as we can
see in Fig.~\ref{fig:per_b}.
In both the case of eccentric and circular orbit, we note that in low dense clusters,
i.e. model A and B, the BHB preserves a pericentre relatively close to the initial value
indicating that such stellar systems are less efficient in favouring the BHB shrinkage.
In such models the pericentres data appears more spread than in models C and D.
A further difference is that for example in model A00, even after 3 Gyr, the BHB have
larger pericentres, indicating that the binary becomes wider, contrary to what is
observed in model A05. We note additionally, that in the intermediate low massive model, B05,
the pericentre reaches very small values (of the order of $10^{-7}\,pc$) which
does not occur for an initial circular orbit.
These results indicate that in such cases, both the cluster stellar 
density and the initial orbital eccentricity play a relevant role in favouring the BHB shrinkage.

\subsection{Retained and dynamically-ejected BHBs}
\label{ret_esc}

We observe that in the majority of cases these dynamically--formed binaries
interacting with other stars in the system can also be ejected away from
the cluster. In the code that we are using, \texttt{NBODY7}, single or binary
stars are considered escapers of their energy is positive and their distance to
the centre of the OC centre is at least two times the initial half mass radius
\citep{aarseth_esc, aarseth_book}.  Taking into account the evolutionary
scenarios discussed in Sect. \ref{BHB_ev}, we derive for each model the
fraction of escaping and retained BHBs.  Table \ref{esc_ret} summarizes the
results of this analysis.

In model A all the BHBs that become harder, i.e. shrink their semi-major axis,
are retained in the OC both in the cases in which the binary has an initial
circular orbit (A00) and in the cases in which the BHBs has initial eccentric
orbit (A05).

In model B only a small fraction of BHBs ($0.7$ \%) is ejected from the cluster
while a large fraction is retained. In particular, in model B05 the fraction of
ejected BHBs ($2.7$ \%) is higher than in model B00. In model C the percentage
of ejected BHBs is larger than in the previous cases. In particular, when the
binary has an initial eccentric orbit (model C05) the fraction of escaping BHBs
is about the $10.5$ \%. Finally, in model D the majority ($\geq 85$ \%) of
BHBs is retained in the cluster even if in this case, contrary to the previous
situations, circular orbits have a higher fraction of ejected BHBs.

After an ionisation of the BHB, the individual black holes usually form a new binary with
a star in the cluster. These dynamically-formed binaries are usually short-lived, and
last at most some tens of Myrs.

After a BHB has been separated, one of the black holes stays in the cluster and
forms a new binary with a star.
These dynamically-formed binaries do not survive for long. Moreover,
we notice that the newly single BHs are more likely to be expelled from the
stellar systems than retained because of multiple scattering with massive stars ($\gtrsim 10$ \Ms).
 The presence of such massive stars is comparable to the time at which the BHs are expelled
 from the systems, which is generally short ($\lesssim 100$ Myr).
 As it is shown in Table \ref{esc_ret}, only in
three models studied (A00, C00 and D00) the BHs are retained after the binary
breaking.

Note that the larger fraction of ejected BHBs "belongs" to more massive
clusters (C and D) in spite of their larger escape velocity.

The time at which the bound BHB is ejected from the cluster varies among the
models, with a BHB mean ejection time between $0.4$ and $1.2$ Gyr. We noticed
that low dense clusters (models A and B) show a BHB ejection time shorter than
massive clusters (models C and D).

\begin{table}
\centering
\caption{
Percentage of BHBs retained (ret) by the cluster or ejected (esc). The first
column indicates the models. Column 2 and 3 indicate the percentage of retained
or ejected BHB that become harder. Column 4 and 5 refers to wider BHBs. Column
6 and 7 give the percentage of retained and ejected single black hole after the
binary breaking.}
\label{esc_ret}
\begin{tabular}{@{}ccccccc@{}}
\toprule
\textbf{Model} & \multicolumn{2}{c}{\textbf{hard}} &\multicolumn{2}{c}{\textbf{wider}} & \multicolumn{2}{c}{\textbf{break}} \\ \midrule
               & ret              & esc&               ret              & esc             & ret               & esc              \\
               & \%               & \%               & \%              & \%              & \%                & \%               \\
A00   & 89.1     & 0.0      & 7.9      & 0.0     & 0.7     & 2.1     \\
A05   & 97.1      & 0.0      & 2.1      & 0.0     & 0.0     & 0.7     \\
B00   & 91.8      & 0.7      & 2.7      & 0.0      & 0.0     & 4.8      \\
B05   & 91.3      & 2.7      & 2.0      & 0.0     & 0.0     & 4.0      \\
C00   & 88.6      & 4.9      & 0.0      & 0.0     & 0.7     & 5.6      \\
C05   & 86.5      & 9.9     & 0.0      & 0.0     & 0.0     & 3.5     \\
D00   & 85.5      & 8.6      & 0.0      & 0.0     & 0.7     & 5.0      \\
D05   & 90.0      & 7.2      & 0.0      & 0.0     & 0.0     & 2.8      \\ \bottomrule
\end{tabular}
\end{table}

\begin{figure*}
  \includegraphics[width=1\textwidth]{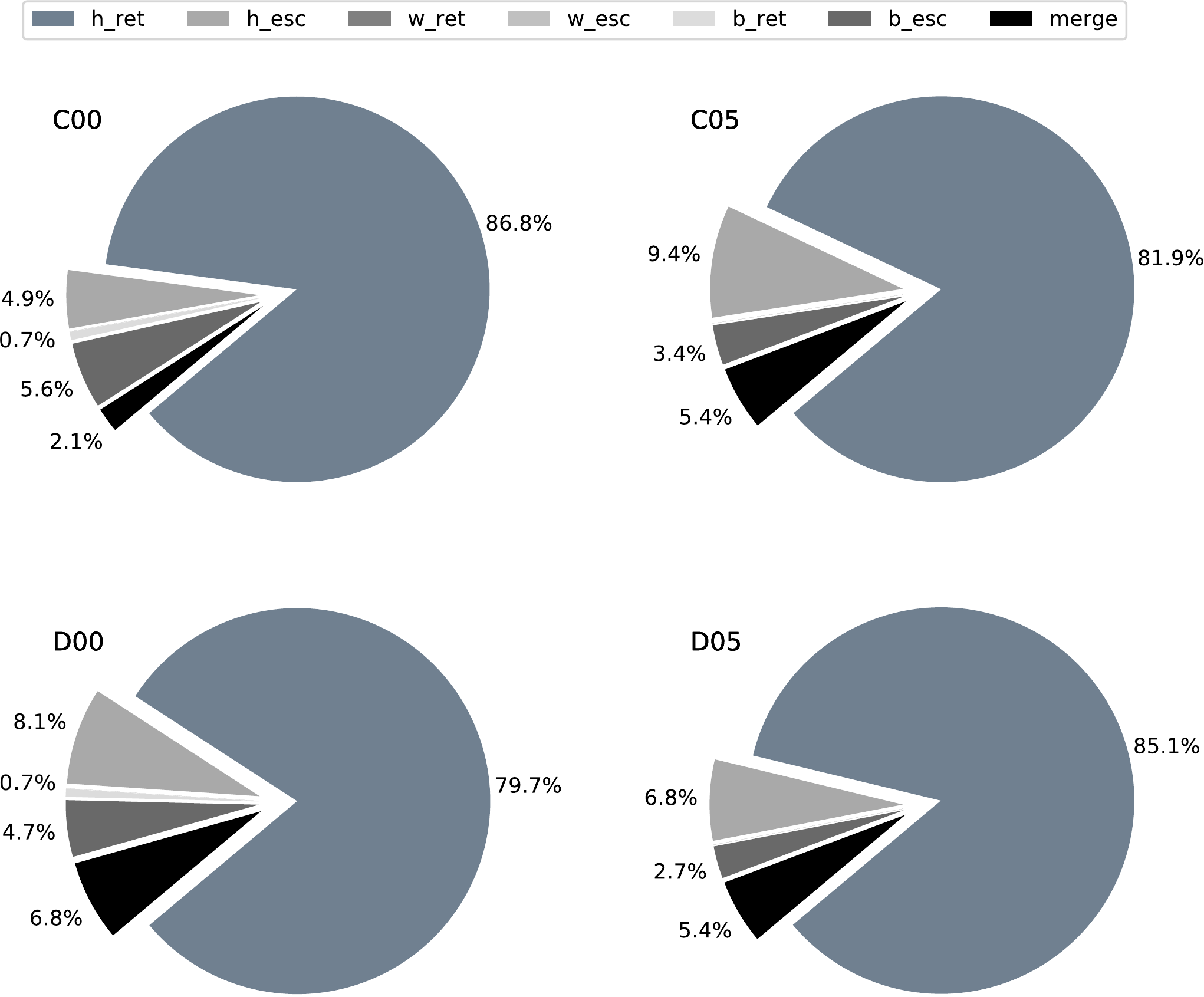}
  \caption{
The pie charts indicate the various evolutionary scenario of the BHB discussed
in Sect. \ref{BHB_ev} and Sect. \ref{ret_esc} for models C and D. The colour is
referred to: the fraction of retained harder BHB ($h_{\rm ret}$), the fraction of
ejected harder BHB ($h_{\rm esc}$), the fraction of escaped wider BHB ($w_{\rm ret}$),
the fraction of retained wider BHB ($w_{\rm ret}$), the fraction of broke binaries
retained ($b_{\rm ret}$), the fraction of broke binaries ejected ($b_{\rm esc}$) and
the fraction of mergers ($merge$). On the other hand the striped slices
referred to BHB that broke up. The width of each slice indicate the percentage
as shown in Table \ref{esc_ret} and Table \ref{table_merge}.
}
  \label{fig:pie}
\end{figure*}

The pie charts in Fig. \ref{fig:pie} illustrate the probabilities of the
different channels for two models studied, C and D (both configuration 00 and
05). Harder binaries are denoted with letter "$h$", wider with "$w$", broken up
binaries with "$b$". Then, each of the three scenarios are split into two
cases: BHB retained by the cluster (indicate with "$ret$") and BHB or BHs
ejected from the system (indicate with "$esc$").  From the pie charts it is
clear that in the majority of cases the BHBs shrink the semi major axis,
becoming harder and remaining bound to the parent cluster. Model C00 and D00
show also a very small fraction of broken up binaries (hence newly single BHs)
which are retained by the clusters. Such result, on the contrary, is not
observed in model C05 and D05. A considerable number of harder BHB escaped from the
cluster is observed in model C05.
 Furthermore, the percentage of newly single BHs escaped from the cluster ($b_{\rm esc}$) is higher
in model C00 and D00. Finally it is worth noticing the fraction of
coalescence events (black slices) in each model.

\subsection{External Tidal Field} For a Milky Way-like galaxy the dynamical
evolution of open clusters may be significantly influenced by an external tidal
field \citep{sambaran1}.  To investigate such effect, we assume our clusters
are embedded in a tidal field like that of the solar neighbourhood.  The
Galactic potential is modelled using a bulge mass of MB = $1.5 \cdot 10^{10}$
\Ms \citep{fdisk} and disc mass MD = $5 \cdot 10^{10}$ \Ms. The geometry of the
disc is modelled following the formulae of \citet{fbulge} with the following
scale parameters a=5.0 kpc and b=0.25 kpc. A logarithmic halo is included such
that the circular velocity is 220 km/s at 8.5 kpc from the Galactic center.
Adopting these configurations, we ran a further sub-set of simulations for each
model A, B, C and D.  The external tidal field generally contribute stripping
stars from the cluster, accelerating its dissolution through the field. In our models
the complete dissolution of the clusters occur between 1.5 Gyr and 3 Gyr.

We notice that the significant reduction of the BHB semi major axis (up to 1-2
order of magnitude) occurs in a time which ranges between $\sim 50$ and $\sim 7
\cdot 10^{2}$ Myr. In such time-range the clusters are still bound and the
tidal forces have not yet contribute to dilute the clusters, avoiding the
binary harden.  The gravitational interactions that contribute to significantly
 shrink the BHB semi major axis act in a short time-range and in such time the cluster
still contain between 60\% and 80\% of bound stars.

The complete disruption of clusters occur when the gravitational interactions
do not play anymore a dynamical role in the evolution of the black hole binary.
It is worth mentioning that such result are typical of open cluster that lie at
8.5 Kpc from the Galactic center, otherwise clusters closer to the central
regions would dissolve in a shorter time scale.

\section{Sources of gravitational waves}

\subsection{Relativistic binaries}
\label{subsec.relbin}

The code that we used for this work (\nbody) identifies those compact objects
that will eventually merge due to the emission of gravitational radiation. Note
that the \nbody~ code indicates a binary as `merging' when at least one of the
conditions described in \cite{aarsethnb7} is satisfied
\footnote{https://www.ast.cam.ac.uk/~sverre/web/pages/pubs.htm} (see also \citet[Sect. 2.3.1]{sambaran3}).
  However, the
code does not integrate in time these binaries down to the actual coalescence,
because this would require a reduction of the time-step down to such small
values to make the integration stall. In Table~\ref{table_merge} we give the
percentage of BHB mergers as identified by \texttt{NBODY7} in the simulations.
The more massive the cluster, the larger the number of relativistic mergers
found. We noted that the initial value of the binary eccentricity is not
necessarily correlated with the number of coalescences. The majority of mergers
occur in a time range between $5$ Myr and $1.5$ Gyr. Only two merger events
take longer, between $\approx 1.5$ Gyr and $\approx 2$ Gyr.
In our models, the clusters have not yet disrupted when the BHB 
coalescences occur, still containing more than the 80 \% of the initial number of stars.

\begin{table}
\centering
\caption{Percentage of BHB mergers found for each model studied.}
\label{table_merge}
\begin{tabular}{@{}cc@{}}

\toprule
\textbf{Model} & \% Mergers
               \\ \midrule
A00            & 0.0                                                                \\
A05            & 0.7                                                                \\
\midrule
B00            & 0.7                                                                \\
B05            & 0.7                                                                \\
\midrule
C00            & 2.1                                                                \\
C05            & 4.3                                                                \\
\midrule
D00            & 7.1                                                                \\
D05            & 5.7                                                                \\ \bottomrule
\end{tabular}
\label{table.mergers}
\end{table}

In Figs.~\ref{fig:parammergea} and \ref{fig:parammergeb} we show the evolution
of the BHB semi-major axis and pericentre distance of a few representative
cases of Table~\ref{table.mergers} which initially were circular or
eccentric, respectively.  It is remarkable that the pericentre distances
drop down to $7--8$ orders of magnitude with respect to the initial value. The
eccentricities fluctuate significantly, episodically reaching values very
close to unity.

\begin{figure}
\centering
\includegraphics[width=0.50\textwidth]{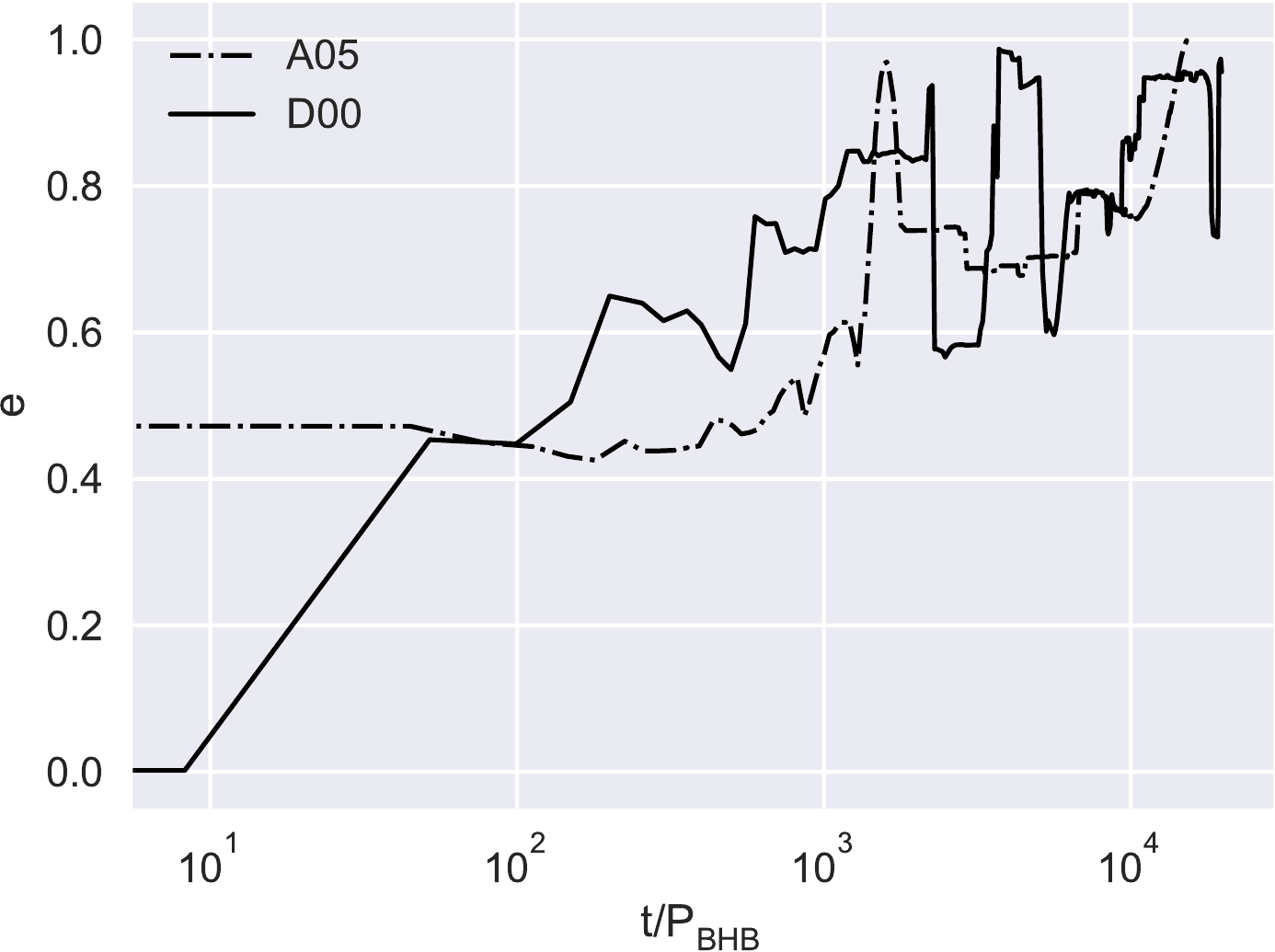}
\caption{The evolution of the eccentricity for two cases in which
the BHBs merge, for models A05 and D00.}
\label{fig:eccmer}
\end{figure}

Because of the relativistic recoil kick
\citep{CampanelliEtAl2007,BakerEtAl2006,GonzalezEtAl2007,fragk17,frlgk18}, the product
of the merger of the BHB might achieve very large velocities, such to escape
the host cluster in all of the cases due to the very small escape velocity.
Fig.~\ref{fig:bhbesctgw} shows the distribution of the BHB semi-major axis
and eccentricity in the last output before the gravitational wave regime drives the merger.

Because these binaries have
undergone many dynamical interactions with other stars, the eccentricities are
very high, ranging between $0.99996$ and above $0.99999$.

Taking into account the expression for the GW emission time, $\mathcal{T}_{gw}$, \citep{peters64},
\begin{equation}
\centering
\mathcal{T}_{gw} (yr) = 5.8 \enskip 10^{6} \enskip \frac{(1+q)^{2}}{q} \enskip \left(\frac{a}{10^{2}\quad \rm{pc}}\right)^{4} \enskip  \left(\frac{m_{1}+m_{2}}{10^{8} \quad \rm{M}_{\odot{}}}\right)^{-3} (1-e^{2})^{7/2}
\label{peters}
\end{equation}
where $q$ is the mass ratio between the two BHs of mass $m_{1}$ and $m_{2}$
\footnote{note that the r.h.s of Eq. \ref{peters} is invariant on the choice $q=m_1/m_2$ or $q=m_2/m_1$},
 we found that about 50\% of the mergers are mediated by a three body encounter with a
 pertuber star.Such three body interaction is thus a fundamental ingredient for BHB coalescence
 in low dense star clusters, as already pointed out by \citet{sambaran3}.
  An example of such mechanism is discussed in the next section (\ref{mikkola}).

\begin{figure} \centering
\includegraphics[width=0.50\textwidth]{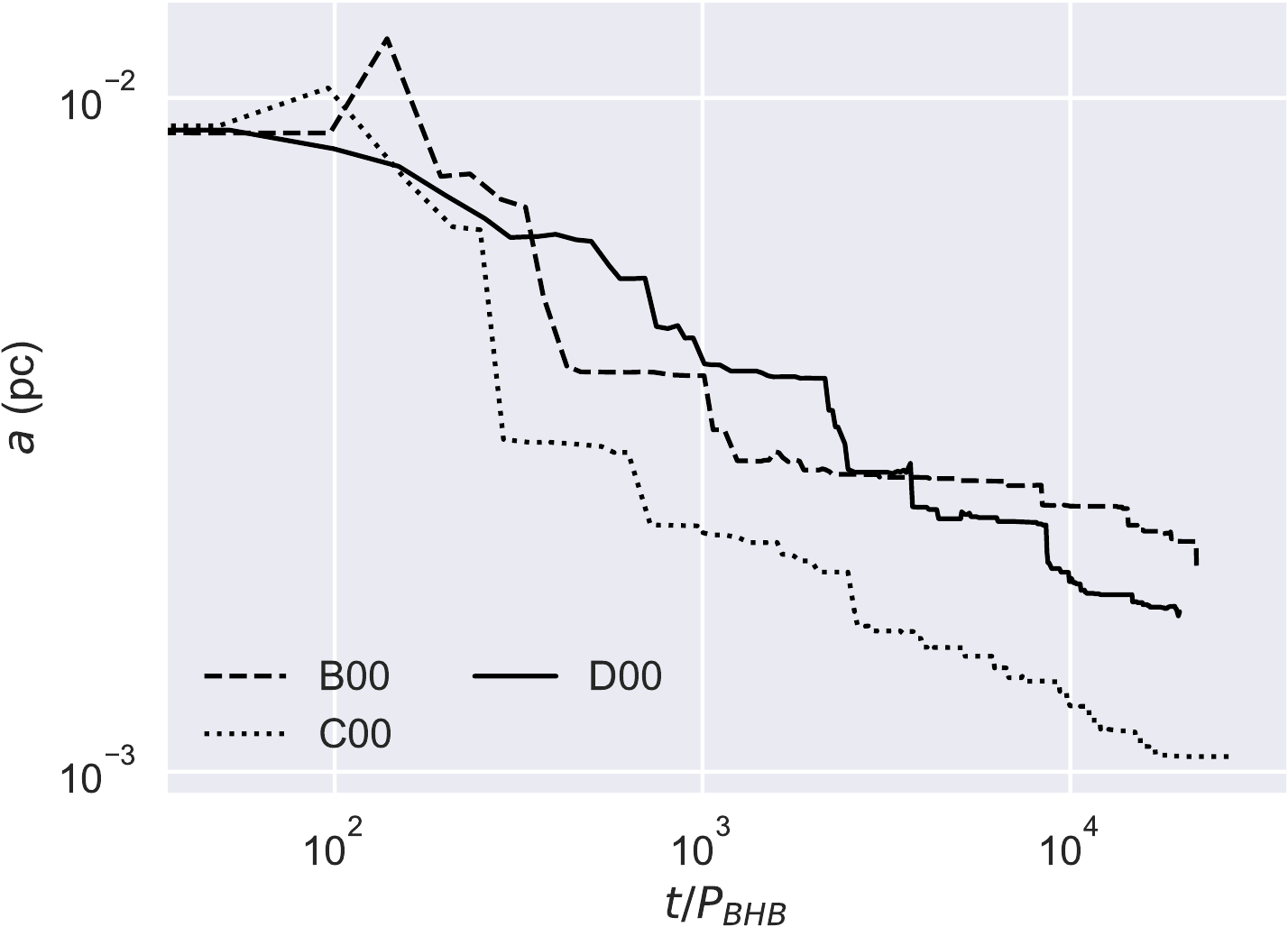}
\includegraphics[width=0.50\textwidth]{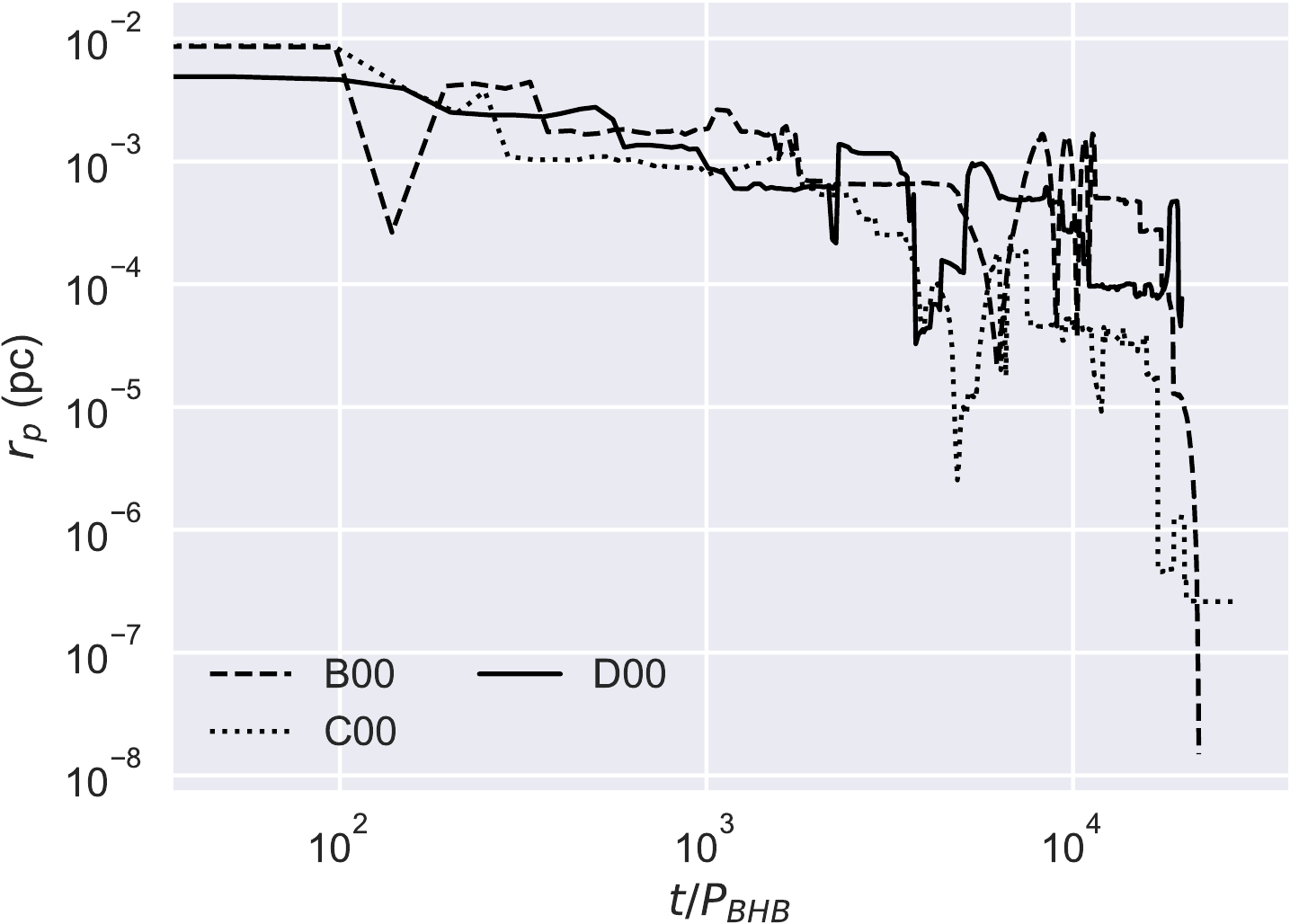}
\caption{Evolution of the BHB semi major axis (upper panel) and pericentre distance
(lower panel) of three illustrative cases which initially were circular.
}
\label{fig:parammergea}
\end{figure}

\begin{figure}
\centering
\includegraphics[width=0.50\textwidth]{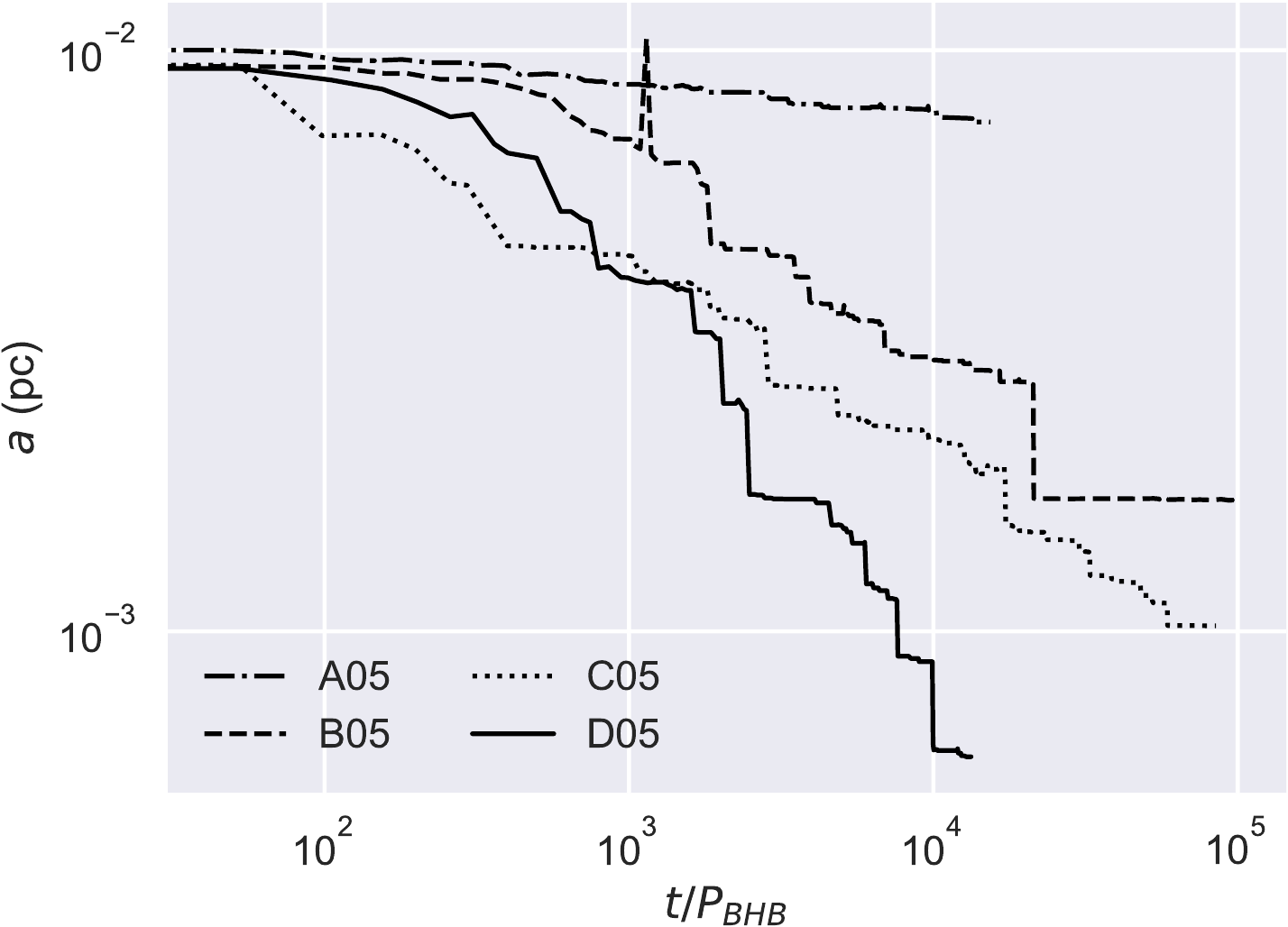}
\includegraphics[width=0.50\textwidth]{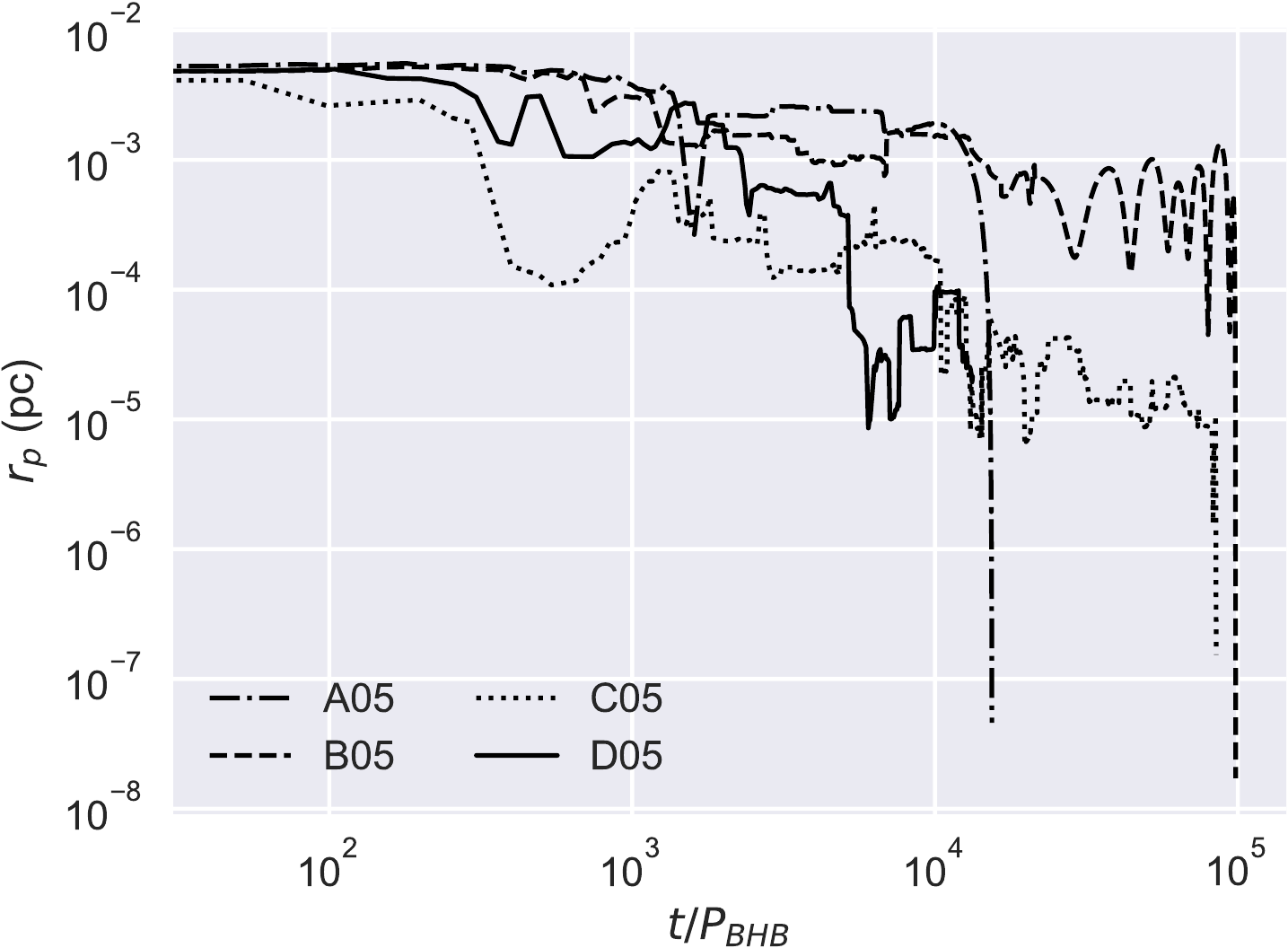}
\caption{Same as Fig.~\ref{fig:parammergea} but for BHBs which initially had an eccentric orbit.}
\label{fig:parammergeb}
\end{figure}

\begin{figure}
\resizebox{\hsize}{!}
          {\includegraphics[scale=1,clip]{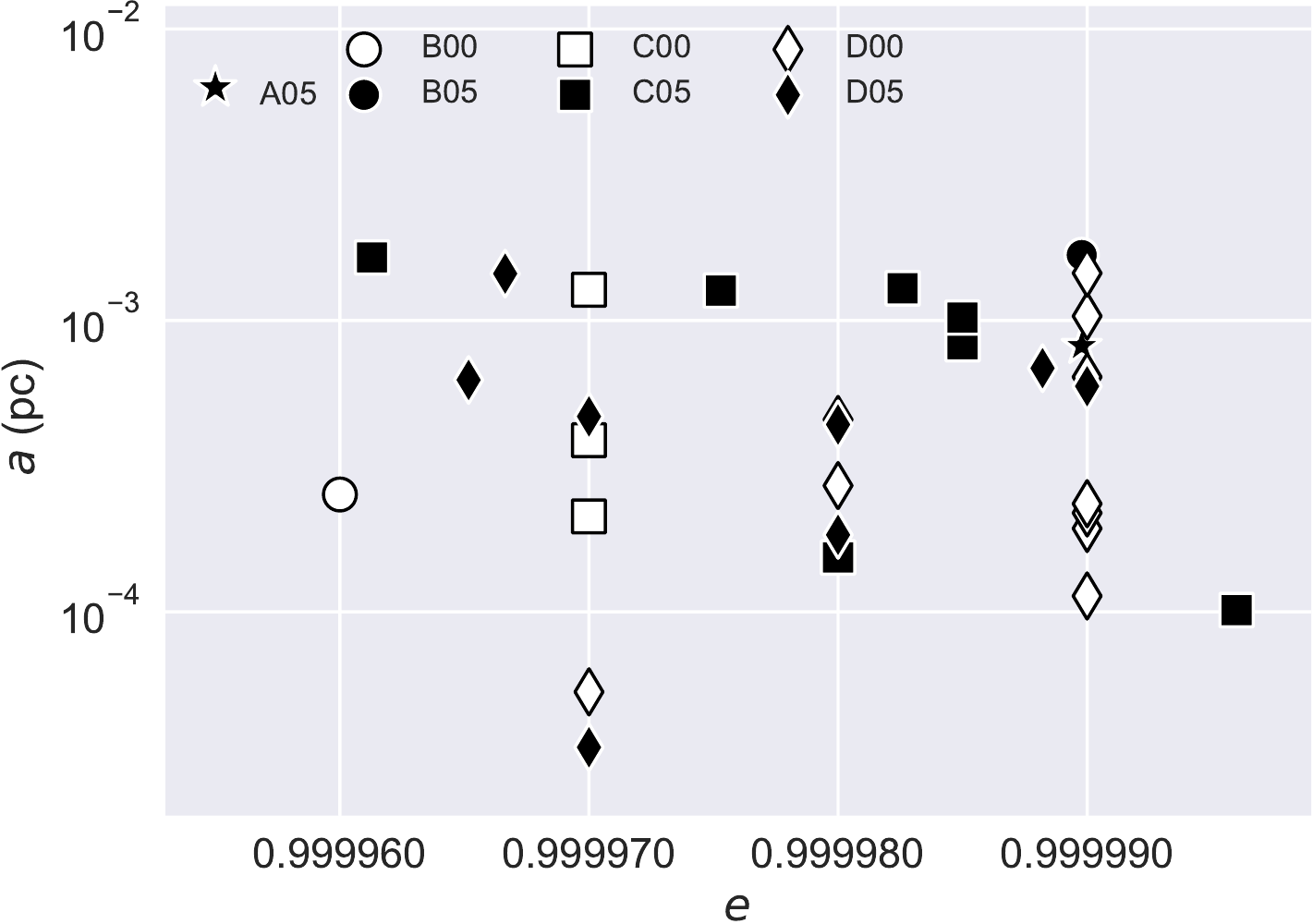}}
\caption
   {
   Distribution of the semi-major axis ($a$) and eccentricity ($e$) for all BHBs
which merge in our simulations. The various symbols refer to the models as defined in Table~\ref{ictab}.
   }
\label{fig:bhbesctgw}
\end{figure}

\subsection{A detailed example of a merger event}
\label{mikkola}
As we said above, \nbody ~identifies the binary merger events in a different
way when a close interaction with a third object occurs.

However, it does not fully integrate those specific cases because following the
detailed binary evolution would practically make the code stuck. So, in order
to check with accuracy the process of BHB coalescence upon perturbation, we
followed the evolution of one of the allegedly merging BHB by mean of the
few-body integrator \argdf \citep{megan17}. Based on the \archain code
\citep{mikkola99}, \argdf includes a treatment of dynamical friction effect in
the algorithmic regularization scheme, which models at high precision strong
gravitational encounters also in a post-Newtonian scheme with terms up to the
$2.5$ order (\citealt{mikkola08}, whose first implementation in a
direct-summation code is in \citealt{KupiEtAl06}).  We chose, at random, one of
our simulations of the D00 model to set initial conditions for the high
precision evolution of a "pre merger" BHB considering its interaction with the
closest 50 neighbours, number that we checked sufficient to give accurate
predictions at regard.

\begin{figure} \resizebox{\hsize}{!}
{\includegraphics[scale=1,clip]{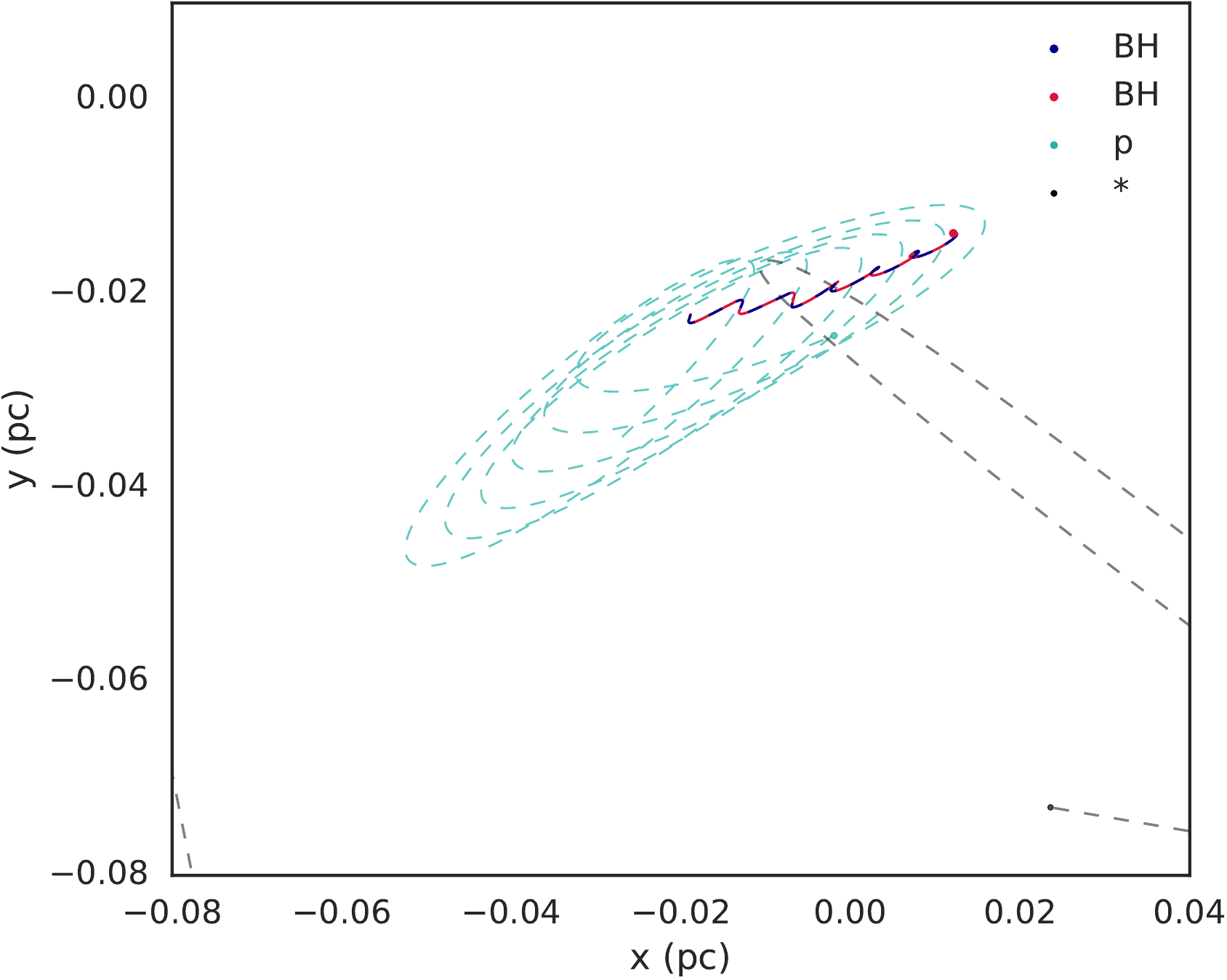}} \caption { Trajectories of the
BHs in our resimulation (model D00). The cyan circle and dashed line represents
the perturbing star and its trajectory, the black holes are shown as a blue 
and red circle and solid lines. The grey circle and lines indicate the stars of the sub
cluster sample simulated. } \label{3bd} \end{figure}

\begin{figure}
\resizebox{\hsize}{!}
          {\includegraphics[scale=1,clip]{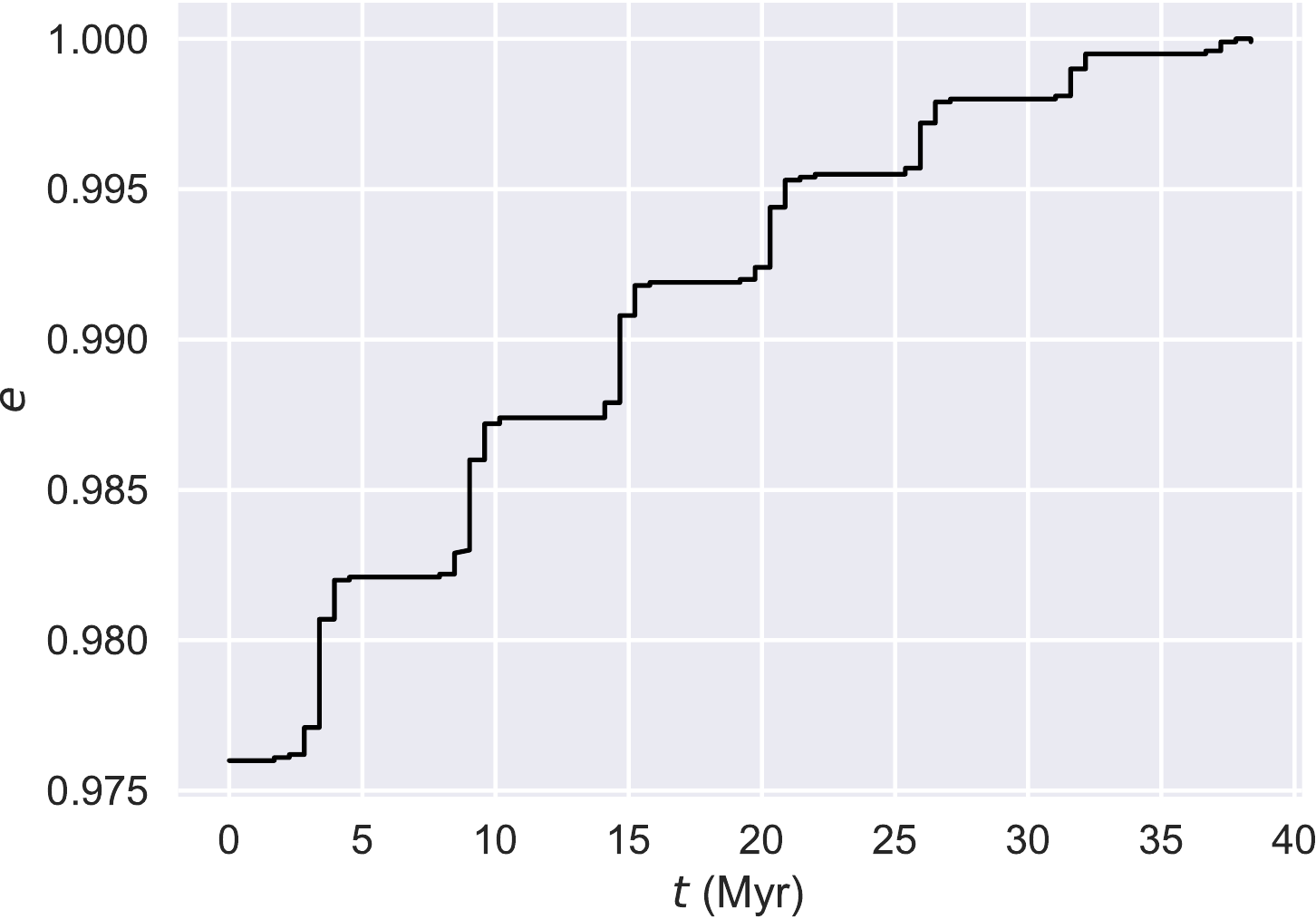}}
\caption
   {
Evolution of the BHB eccentricity of Fig.~\ref{mkk} as a consequence of the
three body encounter. Each jump in the eccentricity corresponds to a close
passage of the third star to the BHB, as described in the text.
   }
\label{mkke}
\end{figure}

\begin{figure}
\resizebox{\hsize}{!}
          {\includegraphics[scale=1,clip]{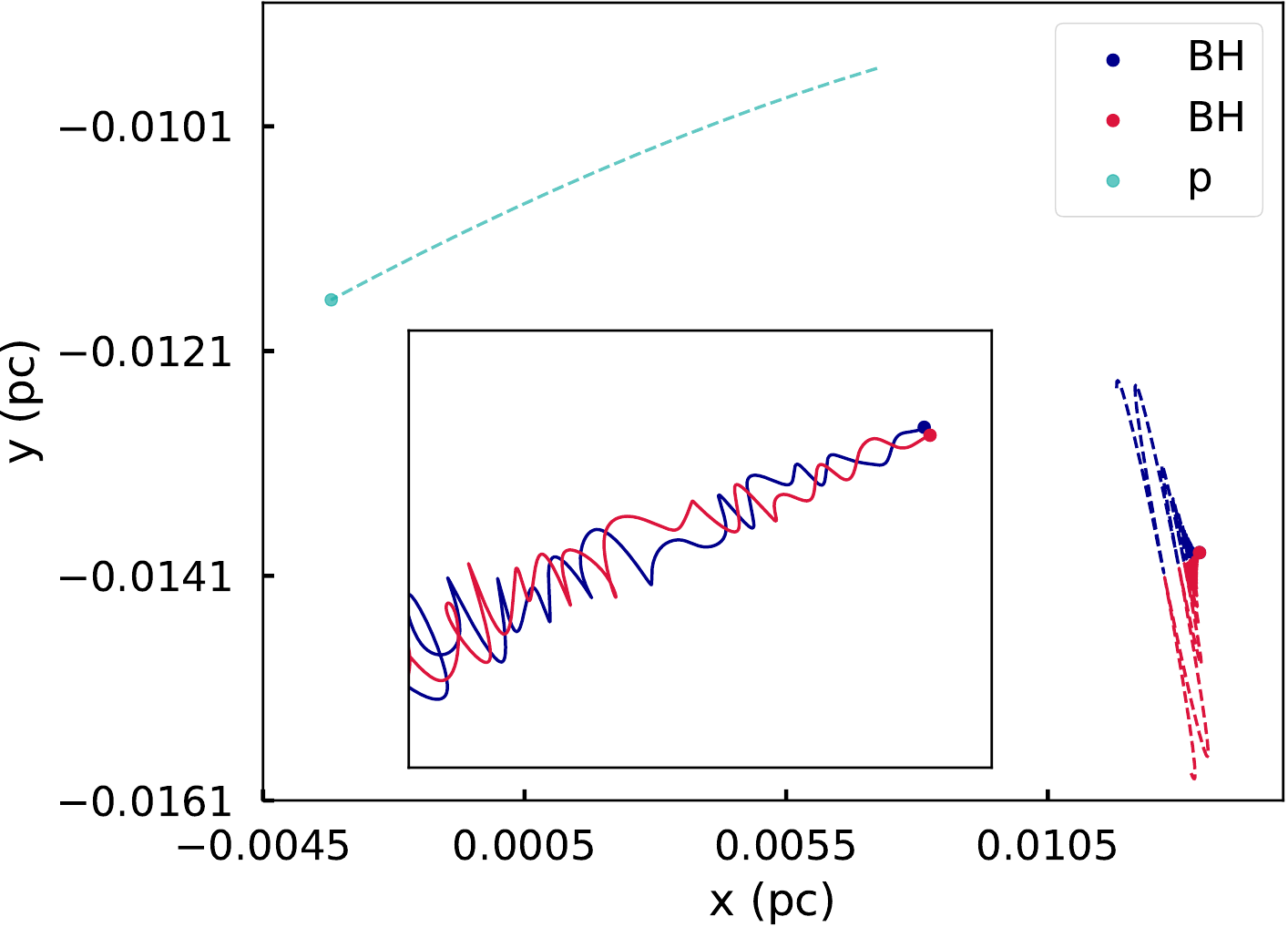}}
\caption
   {
Trajectories of the BHs in our resimulation (model D00).
The cyan circle and dashed line represents the perturbing star and its
trajectory, the black holes are shown as a blue and red circle and solid lines.}
\label{mkk}
\end{figure}

This integration is a clear example of the relevance of dynamical interactions
with other stars. Fig.~\ref{3bd} is a snapshot of the BHB evolution and the
formation of a triple system with a pertuber star. \footnote{An animation of
the triple orbit and the eccentricity evolution is available on line with the name 
triple\_argdf.avi}. The BHB shrinks by
interacting with such pertuber, of mass $3.4\,M_\odot$, which is in retrograde
orbit as compared to the inner binary with an inclination of $105\degree$ indicating
an eccentric \citet{kozai} \citet{lidov} mechanism \citet{naoz}.
 We note also a flyby star of mass
$0.5\,M_\odot$ which interacts with the triple system (BHB \& pertuber) In
Fig.~\ref{mkke} we display the step-like increase of the BHB eccentricity,
which is marked by the repeated interactions with the outer star. Each time the
pertuber orbits around the BHB we observe a step increasing of the
eccentricity. On the contrary the flyby encounter is not efficient to make a
significant perturbation on the eccentricity evolution.  Fig.~\ref{mkk} shows a
zoom of the evolution of the BHB latest orbits before the coalescence event.
The plot in the rectangle is a zoom of the final part of the BHB trajectory (at
its right side), spanning a length scale $\sim 10^{-7}$pc.
Therefore, in this particular case the triple built up is the main ingredient that drives 
the BHB coalescence. A similar result is derived by \citet{sambaran3} for 
low dense star clusters-like.

\subsection{Gravitational Waves}

\begin{figure}
   \includegraphics[width=1\columnwidth]{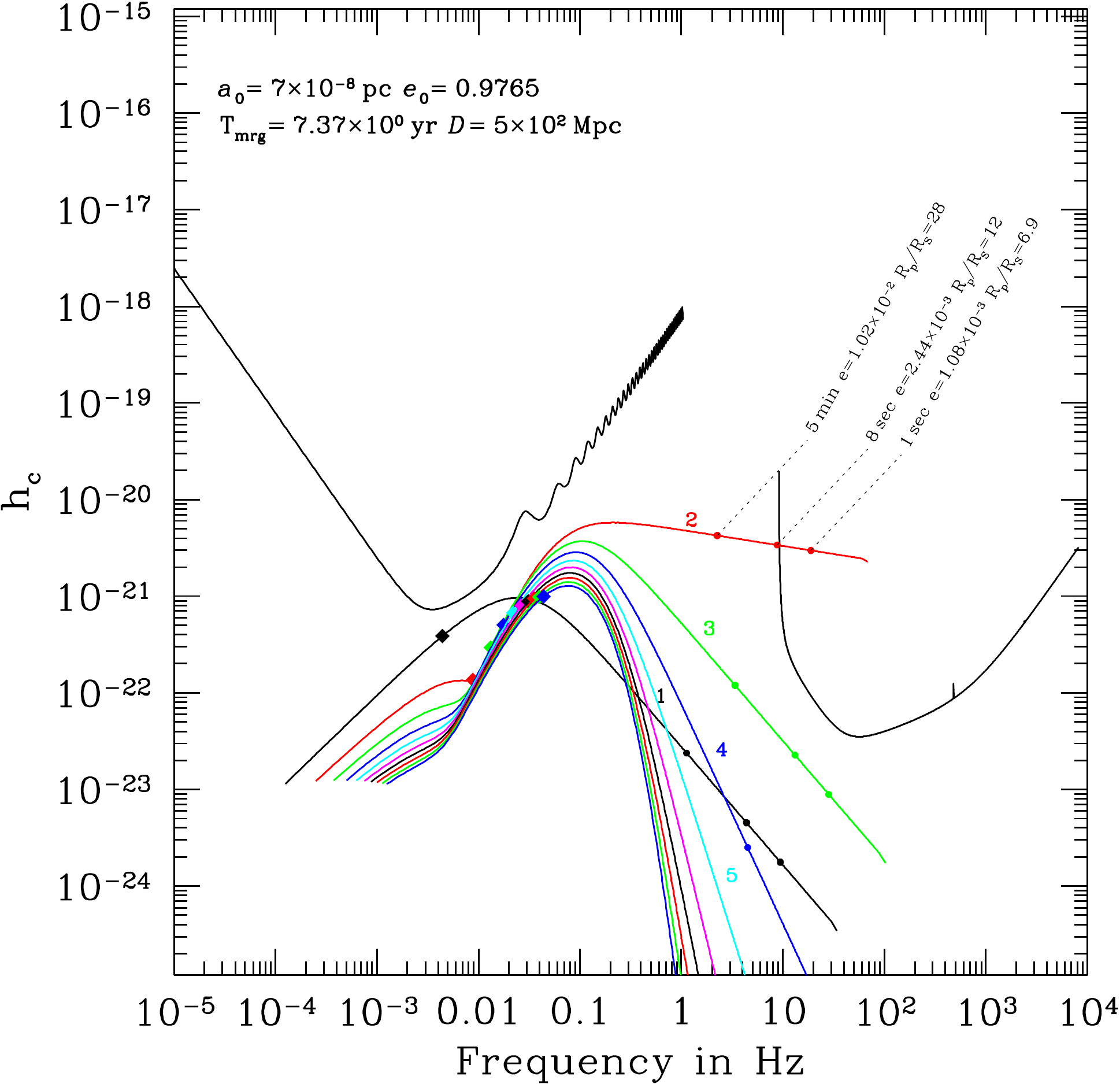}
   \caption{
Characteristic amplitude $h_c$ of the first most important harmonics for the model of
 Fig.~\ref{mkk} at a luminosity distance of $D=500~{\rm Mpc}$. We also
 pinpoint seven moments in the evolution with dots which correspond,
 respectively, to 5 min, 8 sec and 1 sec before the merger of the two black
 holes.
} 
   \label{fig.30_30_Peters}
 \end{figure}

In Fig.~\ref{fig.30_30_Peters} we show the amplitude vs frequency of emitted
gravitational waves for the case described in the above subsection.  Using the
last orbital parameters of the binary which correspond to the last integration
made with \argdf, we evolve the last phase of the binary by means of
Eq.\ref{peters} deriving a coalescence time $T_{\rm mrg} \cong
7\,\textrm{yrs}$.  The amplitude is estimated following the approach of
Keplerian orbits of \cite{PM63} and the orbital evolution as in the work of
\cite{peters64}. We have set the luminosity distance to that of the first
source detected by LIGO \citep{abbott16a}, which corresponds to a redshift of
about $z=0.11$. As described by the work of \cite{ChenAmaro-Seoane2017}, only
circular sources are audible by LISA, which is ``deaf'' to eccentric binaries
of stellar-mass black holes that emit their maximum power at frequencies farther
away from LISA. Hence, this particular source only enters the
Advanced LIGO detection band.

\subsection{Black holes inspiraling outside of the cluster}

In our simulations some BHBs undergo a strong interaction with a star and they
are kicked out from the cluster. The BHBs become escapers as defined in
Section~(\ref{ret_esc}). In this case, the BHBs remain almost frozen in their
relative configuration without any possible further evolution of their orbital
parameters as described in Section~(\ref{subsec.relbin}): the escaping BHB
evolves only due to the emission of gravitational radiation. For all these
escaping BHBs (47 cases over the whole set of our simulations), we estimate the
timescale for coalescence using the approach of Keplerian orbits of
\cite{peters64} and find that it always exceeds the Hubble time.

The inspiral phase of these binaries falls in the sensitivity window of LISA.
However, they evolve very slowly in frequency due to the fact that the
semi-major axis is still large, and the time to coalescence scales as $\propto
a^4$. For an observational time of 5 years, the source would virtually not have
moved in frequency, and hence the accumulated SNR over that time is negligible.

\subsection{Merger Rate}

To estimate approximately the merger rate, $\mathcal{R}_{\rm mrg}$
we first derive the mean number density of open clusters, $n_{\rm OC}$,
over a volume $\Omega$ corresponding to redshift $z\leq 1$ as

\begin{equation}
{n_{\rm OC}}= \frac{{N_{\rm OC-MW}} \enskip {N_{\rm MW-\Omega}}}{\Omega}.
\end{equation}

\noindent
In this equation ${N_{\rm OC-MW}}$ is the number of OCs in Milky Way (MW)-like
galaxies and  ${N_{\rm MW-\Omega}}$ is the number of MW-like galaxies within
${z=1}$. We estimate the number of OCs in our Galaxy on the basis of the
open-cluster mass function discussed in \cite{piskunov08,spz2010} for the mass
range of OCs considered in our work (from $300$ \Ms to approximately $3000$
\Ms).  We take N$_{\rm {MW}}=10^8$ as the number of Milky Way-like galaxies at
redshift $\sim 1$, as discussed in \cite{tal17}.
We stress here that the estimated merger rate is an upper limit, since it assumes 
that each open cluster host a massive BHB similarly to the clusters studied in our models.

Hence, the black hole binary meger rate can be estimated to be

\begin{equation}
\centering
\mathcal{R}_{\rm mrg} = \frac{1}{N_{\rm s}} \sum_{k=1}^{N_{\rm s}} \frac{n_{\rm OC}}{t_{\rm k}} \approx 2 \textrm{Gpc}^{-3}\,\textrm{yr}^{-1},
\label{rate}
\end{equation}

\noindent
where N$_{\rm s}$ is the total number of $N$-body simulations performed in this
work, and t$_{\rm k}$ is the time of each coalescence event as found in our
simulations.  This estimate is however derived under the most favourable
conditions and represents the most optimistic merger rate expected from
low-mass open clusters. Note that the BHB merger rate inferred from the first
LIGO observations (GW150914) is in the range $2$ - $600$ Gpc$^{-3}$ yr$^{-1}$
\citep{abbrate16}. The most updated estimate of the merger 
rate from LIGO-Virgo events (after including GW170104) is $12$-$213$ Gpc$^{-3}$ yr$^{-1}$ \citep{abbott17}.
Our BHB merger rate is consistent with those found in
\cite{sambaran1,sambaran2,sambaran3} for BHB mergers in Young Massive Star Clusters
(YMSC).  \cite{antoninirasio2016} found a merger rate ranging from $0.05$ to
$1$ Gpc$^{-3}$ yr$^{-1}$ for possible progenitor of GW150914 in globular
clusters at z$<$0.3. \cite{rodriguez16b} and \cite{rodriguez16} derived that in
the local universe BHBs formed in Globular Clusters (GCs) will merge at a rate of $5$ Gpc$^{-3}$
yr$^{-1}$.  A result very similar was derived by \citet{askar17} who, for BHB
originated in globular cluster, derived a rate of $5.4$ Gpc$^{-3}$ yr$^{-1}$. 
When the history of star clusters across cosmic time is included, \citet{frak18} 
showed that the rate in the local Universe is $\sim 10$ Gpc$^{-3}$, i.e. nearly twice the rate predicted for isolated clusters.

\begin{figure} \centering \includegraphics[width=0.50\textwidth]{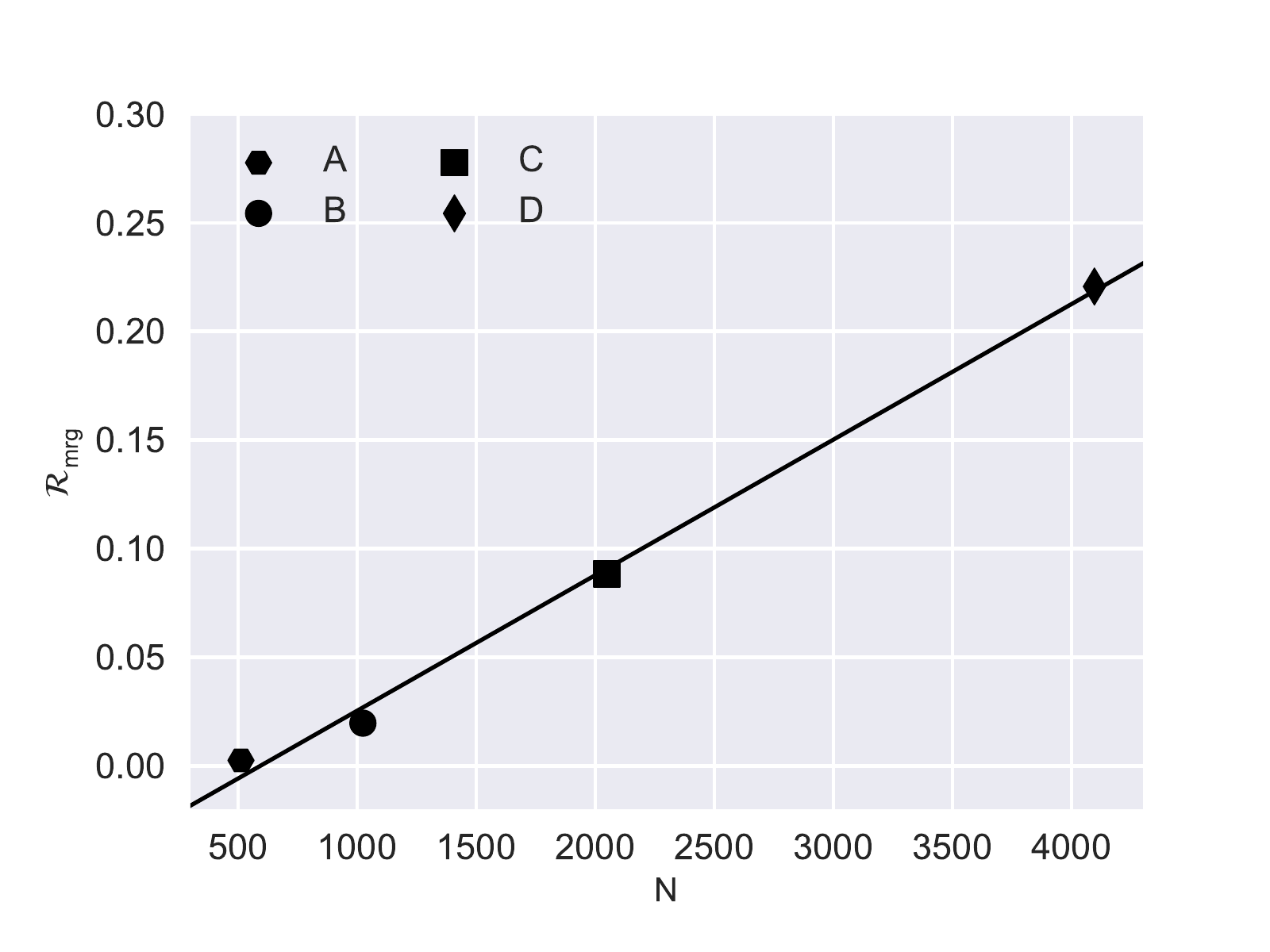}
\caption{The most optimistic merger rate ($\mathcal{R}_{\rm mrg}$, in Gpc$^{-3}$ yr$^{-1}$)
obtained for each model studied in this work, as function of the total initial number of cluster stars N.
The merger rates are well fitted with a linear relation, $R_{\rm mrg}=aN+b$, where
$a= 6.2e-{05}$ and $b= -0.04$.}  
\label{fig:mrate}
\end{figure}

In Fig.~\ref{fig:mrate} we show the estimated merger rate as a function of
the initial number of cluster stars ($N$). The merger rates derived from our
models A, B, C and D are well fitted with a linear relation.
An extension of our merger rate estimate to globular cluster-like systems ($N >
10 ^{5}$) gives a result in agreement with that found in \cite{park17}, and
previously found by \cite{bae14} and \cite{rodriguez16,rodriguez16b}.

Although BHB mergers originating in open clusters-like systems might be less
numerous than those produced in massive star clusters, they would add a
comparable amount to the BHB merger rate in the Universe because of their 
larger abundance \citep{sambaran2,sambaran3}.

\section{Tidal disruption events and BHB ejection}
\label{tde}

All numerical models we have considered so far have solar metallicity, $Z =
2.02$, and are based on the standard stellar evolution recipes
\citep{hurley2000, hurley2002}.

Moreover, they all consider an equal-mass BHB sitting in the host cluster
centre with an initial mass $M_{\bhb} = 60$ \Ms.  In order to explore the role
played by metallicity, stellar evolution recipes adopted, BHB mass and mass
ratio, we present in this section an additional sample consisting of 421
simulations (the supplementary models), gathered in 5 different groups.

In all these 5 supplementary groups, the OC initial conditions are the same as
in D00 principal models. This implies $N_{\rm cl} = 4096$ and, for the BHB
initial orbit, $a_\bhb = 0.01$ pc and $e_\bhb = 0$, unless specified otherwise.
We labels each group with the letter M and a number between 1 and 5.

In model M1, we model the OC assuming an initial metallicity value
$Z=0.0004$, typical of an old stellar population.  The BHB initial conditions
are the same as in model D00. Stellar evolution is treated taking advantage of
a new implementation of the \texttt{SSE} and \texttt{BSE} tools, which includes
metal-dependent stellar winds formulae and improvements described in
\cite{belczynski2010}. In the following, we identify the updated stellar
evolution treatment used for model M1 as BSEB, while in all the other cases we
label them with BSE. Note that these updates allow the formation of BHs
with natal masses above $30$ \Ms, while this is not possible in the standard
\texttt{SSE} implementation \citep{hurley2000}. Moreover, it must be stressed
that the updates affect only metallicities below the solar value.

Model M2 is similar to model M1 in terms of initial metallicity and BHB initial
condition, while we used the standard \texttt{SSE} and \texttt{BSE} codes to
model stellar evolution.  Therefore, the underlying difference between this and
the previous is that in the latter the mass of compact remnants is
systematically lower. This, in turn, implies that the number of perturbers that
can have a potentially disruptive effect on the BHB evolution is reduced in
model M2.

In model M3 we adopt $Z = 0.02$, i.e. solar values, and we focuse on a BHB with
component masses $M_1 = 13$ \Ms and $M_2 = 7$ \Ms. This set has a twofold
focus. On one side, it allows us to investigate the evolution of a BHB with
mass ratio lower than 1. On the other side, since in this case the BHB total
mass is comparable to the maximum mass of compact remnants allowed from stellar
evolution recipes, gravitational encounters should be more important in the BHB
evolution.

To further investigate the role of mass ratio, M4 models are similar to M3, but
in this case the BHB mass ratio is smaller, namely $q=0.23$, i.e. the
components mass are $M_1 = 30$ \Ms and $M_2 = 7$ \Ms.

In all the principal and supplementary models discussed above, we assume that
the BHB is initially at the centre of the OC. In order to investigate whether
such a system can be formed dynamically, i.e. via strong encounters, in model
M5 we set two BHs, with masses $M_1 = M_2 = 30$ \Ms, initially unbound. In this
case we set $Z=0.02$, in order to compare with D00 principal models.

The results of these runs are summarized in Table \ref{newsim}.

\begin{table}
\centering
\caption{Supplementary models. The columns refer to: model name, BHB individual masses and mass ratio, metallicity, stellar evolution recipes used, number of simulations performed. The cluster is always simulated with 4096 stars.}
\begin{tabular}{ccccccc}
\hline
Model  & $M_1$ & $M_2$ & $q$ & $Z$ & SE & $N_{\rm mod}$ \\
 &  \Ms & \Ms &  & \Zs & &\\
\hline
M1  & 30 & 30 & 1 & $10^{-4}$ & BSEB & 109\\
M2  & 30 & 30 & 1 & $10^{-4}$ & BSE  & 131\\
M3  & 13 & 7 & 0.54  & $1$    & BSE  & 100\\
M4  & 30 & 7 & 0.23  & $1$    & BSE  & 42 \\
M5  & 30 & 30 & 1 & $1$ & BSE  & 89 \\
\hline
\end{tabular}
\label{newsim}
\end{table}

Since we are interested only in the evolution of the initial BHB, we stop the
simulations if at least one of the BHB initial components is ejected away from
the parent OC.

When metallicity-dependent stellar winds are taken into account (model M1), the
reduced mass loss causes the formation of heavier compact remnants, with masses
comparable to the BHB components. Since the number of BHs is $\sim
10^{-3}N_{\rm cl}$, according to a Kroupa IMF, in models M1 at least
4-5 heavy BHs can form, interact and possibly disrupt the initial BHB. This is
confirmed in the simulations results - we find one of the BHB components
kicked out in $P_{\rm esc} = 34.9\%$ of the cases investigated.
After the component ejection, the remaining BH can form a new
binary with one of the perturbers, or a new BHB.

The ``ejection probability'' in models M2 is only slightly lower than
in M1, $P_{\rm esc} = 33.6\%$, thus implying that the heavier perturbers
forming in models M1 only marginally affect the BHB evolution. This is likely
due to two factors: (i) their number is relatively low (4-5), (ii) the mass
segregation process in such a low-density, relatively light stellar system is
slower than the time over which stellar encounters determine the BHB evolution.
The latter point implies that the BHB evolution is mostly driven by the
cumulative effects of multiple stellar encounters, rather than to a few
interactions with a heavy perturber.

In model M3, characterized by a lighter BHB and solar metallicity, the BHB
total mass falls in the high-end of the BH mass spectrum, $20$ M$_\odot$. This
implies a larger number of massive perturbers with respect to the standard case
discussed in the previous sections and provides insight on the fate of light
BHBs in OCs. Due to the high-efficiency of strong interactions, the BHB unbinds
in $f_{\rm esc} = 32\%$ of the cases, and in no case the BHB undergoes
coalescence.

Model M5 is characterized by a similar ejection probability, which instead
rises up to $40.5\%$ in model M4.  This is likely due to the relatively
low-mass of the secondary component. Indeed, as shown through scattering
experiments, BHB-BH interactions seem to naturally lead to a final state in
which the resulting BHB has a larger mass ratio \citep[see for
instance][]{ASKL18}.

In a few cases, we found that the BHB disruption is mediated by a star, which
binds to one of the two BHB former components.  The newly formed BH-star pair
is characterized by a high eccentricity ($e>0.9$) and pericentre sufficiently
small to rip the star apart and give rise to a tidal disruption event (TDE). In
the current \texttt{Nbody6} implementation, only the $10\%$ of the star mass is
accreted on the BH, while this percentage can be as high as $50\%$.

The fraction of models in which a TDE takes place spans one order of magnitude,
being $f_{\rm TDE} \cong 0.03-0.3$, with the maximum achieved in models M4 and
the minimum in M1. Note that in model M5 we did not found any TDE (see Table
\ref{newsim}), but in this case the two BHs are initially moving outside the OC
inner regions.

In our models, TDEs involve either main sequence stars (MS), stars in the core
He burning phase (HB) or in the early asymptotic giant branch (AGB) phase.  In
model M3 ($f_{\rm TDE} = 0.14$) TDEs involve MS ($29\%$), early AGB ($57\%$)
and AGB ($14\%$) stars.  In model M4, where the BHB has a low mass ratio
($q=7/30$), TDEs are boosted, since in this case is easier to replace the
lighter BH. Indeed, a component swap occurs in $28.5\%$ of the cases, with the
new companion star being swallowed by the heavier BH.

Our findings suggest that X-ray or UV emission from OCs can be the signature of
the presence of BHs with masses as high as $20-30$ \Ms.

\begin{table}
\centering
\caption{Summary of results from the supplementary models.  Columns refer to: model name, percentage of cases in which at least one of the BHB components is ejected, percentage of cases in which a star is swallowed by one of the two BHs, percentage of cases in which the BHB merges.}
\begin{tabular}{ccccccc}
\hline
Model & $P_{\rm esc}$ & $P_{\rm TDE}$ & $P_{\rm mer}$ \\
 & $\%$ & $\%$ & $\%$ \\
\hline
M1 & 34.9& 2.8 & 0.0\\
M2 & 33.6& 6.9 & 3.8\\
M3 & 32.0& 14.0& 0.0\\
M4 & 40.5& 28.5& 0.0\\
M5 & 32.6& 0.0 & 0.0\\
\hline
\end{tabular}
\label{newsim}
\end{table}

Using our results we can calculate the TDE rate for Milky Way - like galaxies
as

\begin{equation}
\Gamma_{\rm TDE} = \frac{f_{\rm TDE}N_{\rm OC}N_{\rm MW}}{\Omega T} = 0.3-3.07\times 10^{-6} {\rm yr}^{-1},
\end{equation}

Our estimates nicely agree with similar TDE-rate calculation provided by
\citet{perets16}, and results in a $\sim 1$ order of magnitude lower than the values calculated for TDEs occurring around supermassive black holes
\citep{fraglei18,stonemetzger,stone17,stone16b,megan1}. Here $f_{\rm TDE}$
is the fraction of TDE inferred from simulation, while we adopt the values
for $N_{\rm OC}, ~ N_{\rm MW}$ and $\Omega$ discussed in the previous section.
Moreover, we assumed $T = 3$ Gyr, i.e. the simulated time.

We apply the same analysis to our principal models and find a TDE rate for
solar-metallicity OCs of $\Gamma_{\rm TDE} = 0.3-3.07\times 10^{-6}
{\rm yr}^{-1}$ for MW-like galaxies in the local Universe.

The BHB coalescence occurs in a few cases $f_{\rm mer} \cong 0.004$, and only in
models M2, where metallicity dependent mass loss is disabled. This suggests
that there exists three different regimes, depending on the perturber maximum
mass $M_p$. If (1) $M_\bhb \gg M_p$, the BHB is much more massive than field stars and
stellar encounters poorly affects its evolution; however, if (2) $M_\bhb \geq
M_p$, a few perturbers have masses comparable to the BHB, and can efficiently
drive it toward coalescence, causing for instance an increase in the BHB
eccentricity or a considerable shrinkage; in case (3) that $M_\bhb =
M_p$, there is at least one perturber with a mass similar, or even larger, than
the BHB. The BHB-perturber interactions causes either the BHB disruption, or
the formation of a new BHB with the perturber replacing the lighter BHB
component.

Note that we cannot exclude that a BHB merge in other models, since we stop
the computation if the original BHB gets disrupted. Hence, we can infer a lower
merger rate for metal poor OCs as follows

\begin{equation}
\mathcal{R}_{\rm mrg} = \frac{f_{\rm mer}N_{\rm MW} N_{\rm OC}}{\Omega T} \simeq 0.26{\rm yr}^{-1}{\rm Gpc}^{-3}.
\end{equation}

These models highlight the importance of stellar evolution in our calculations,
since stronger stellar winds lead to smaller remnants reducing the number of
objects massive enough to cause the BHB disruption. This leads to a higher
probability for the BHB to shrink by the repeated interactions with smaller
objects.

As described above, in model M5 the two BHs are initially unbound, and their
initial position and velocities are kept coherently to the OC distribution
function.  In this situation, the fraction of cases in which at least one of
the BHs is ejected from the cluster is similar to that of the other models
($f_{\rm esc}\sim 32.6\%$), but in none of the models the two BHs bind
together.  This is due to the low efficiency of dynamical friction in the OC
that avoids the two BHB to decay in the innermost potential well. Also TDEs are
suppressed, due to the low number of strong encounters between BH and cluster
stars because of the low density of the surrounding environment.

To conclude, our supplementary models confirm that the possibility for a BHB to
coalesce in an OC depends strongly on the environment in which the BHB formed
and on its total mass and mass ratio.  In metal-poor OCs (metal-dependent)
stellar winds drive the formation of a seizable number of massive perturbers
that can efficiently disrupt the BHB, thus reducing the coalescence
probability. Coalescence is strongly reduced also in the case of low mass
ratios ($q\sim 0.2$) or relatively light BHBs ($M_1+M_2 \sim 20$ M$_\odot$).

One of the most interesting outcomes of the models presented in this section is
the possibility to use the OC TDE rate as a proxy to infer the presence of a
massive BH or BHB around the OC centre.

\section{Conclusions}

In this paper we address the evolution of an equal mass, non-spinning, stellar
BHB with total mass $60$ M$_{\odot}$ inhabiting the centre of a
small/intermediate star cluster (open cluster like, OC), using  the direct
$N$-body code \nbody \citep{aarsethnb7}. In order to quantify the effect of
repeated stellar encounters on the BHB evolution, we vary the OC total mass and
the BHB orbital properties, providing a total of $\sim1150$ simulations which
we refer to as {\it principal} models. For the sake of comparison, we also
investigate the role played by the BHB total mass, the stellar evolution
recipes adopted and the OC metallicity. These can be considered as {\it
supplementary} models.  The total simulations sample hence consists of $\sim
1500$ different OC models, with masses in the range $300-3000$ \Ms.

In $\sim 95\%$ of all the principal simulations performed, the BHB hardens due
to the repeated scatterings with flyby stars, while its eccentricity increases
significantly. This process takes place on a relatively short time-scale, $\sim
1$ Gyr. In $\sim 1.2\%$ of the principal simulations, instead, the
perturbations induced by massive stars that occasionally approach the BHB make
it wider.  In the remaining $\sim 4.8\%$ cases, the interactions with OC stars
are sufficiently strong to break up the BHB.  When the BHB gets harder, its semi-major axis reduces by 2 to 4 orders of magnitude, thus
decreasing the merger time-scale by a factor 16 in the best case. Hardened BHBs
are retained within the parent OC with a probability of $95\%$, while those
becoming wider are all retained.  In the case of BHB breakup, the two BHs tend
to form short-lived binary systems with other OC stars, and eventually at least
one of the two BHs is ejected from the parent cluster.

In $\sim 3\%$ of the models, the star-BHB interactions are sufficiently
effective to drive the BHB coalescence within a Hubble time.  We find that a
crucial ingredient for the BHB to merge is the interaction with a perturbing
star, which considerably shortens the merger time. These dynamical perturbers
enhance the number of GW sources by as much as $50\%$.  The merger takes place
in a time ranging from $5$ Myr to $2.9$ Gyr. In a few cases, the merging binaries emit
GWs shifting from the $10^{-3}$ to the $10$ Hz frequency band. This suggests that
merging BHBs in OCs can potentially be seen both by LISA, $\sim 200$ yr before
the merger, and LIGO, during the last phase preceding the merger.

Extrapolating our results to the typical population of OCs in MW-like galaxies
in the local Universe, we found that the most optimistic merger rate for BHB
mergers in low-mass stellar systems is $\mathcal{R}_{\rm mrg}\sim$ 2 yr$^{-1}$
Gpc$^{-3}$, a value compatible with the merger rate expected for galactic
nuclei, but smaller than the merger rate inferred for globular and young
massive clusters.

According to our supplementary models, in low-metal environments the BHB
hardening is suppressed, due to the presence of a large number of high-mass
perturbers that can efficiently drive the BHB disruption. In this regard,
different stellar evolution recipes may affect significantly the results, since
they regulate the maximum mass of compact remnant. Assuming a smaller BHB and a
solar metallicity for the cluster stars leads to similar results, since, again,
the fraction of perturbers sufficiently massive to drive the BHB disruption is
much larger.

In none of the cases in which the BHB components are initially kept unbound the
BHB forms via dynamical processes. This is due to the low efficiency of
dynamical friction in the OC low-dense environment, which is unable to drive
the BHs orbital segregation and pairing. So binaries as the ones considered in
this paper should be primordial.

In a noticeable fraction of the supplementary models, we found that the BHB
breaks up and one of the BHs forms a very eccentric binary with an OC star,
typically a main sequence or an AGB star. These binaries are usually
short-living systems and result in a tidal disruption event, with part of the
stellar debris being swallowed by the BH.

Our supplementary models suggest that TDEs in OCs occur at a rate $\Gamma_{\rm
TDE} = 3.08\times 10^{-6}$ yr$^{-1}$ per MW-like galaxies in the local
Universe.

\section*{Acknowledgements}

SR acknowledges Sapienza, Universit\'a di Roma, which funded the research
project ''Black holes and Star clusters over mass and spatial scale'' via the
grant AR11715C7F89F177.  SR is thankful to Sverre Aarseth of the Institute of
Astronomy, Cambridge, for his helpful comments and suggestions during the
development of this work.  MAS acknowledges the Sonderforschungsbereich SFB 881
"The Milky Way System" (subproject Z2) of the German Research Foundation (DFG)
for the financial support provided. PAS acknowledges support from the Ram{\'o}n
y Cajal Programme of the Ministry of Economy, Industry and Competitiveness of
Spain, as well as the COST Action GWverse CA16104. GF is supported by the Foreign Postdoctoral Fellowship Program of the Israel Academy of Sciences and Humanities. GF also acknowledges support from an Arskin postdoctoral fellowship and Lady Davis Fellowship Trust at the Hebrew University of Jerusalem. 

\bibliographystyle{mnras}
\bibliography{rasetal}

\newcommand{\noop}[1]{}
\begin{thebibliography}{}
\makeatletter
\relax
\def\mn@urlcharsother{\let\do\@makeother \do\$\do\&\do\#\do\^\do\_\do\%\do\~}
\def\mn@doi{\begingroup\mn@urlcharsother \@ifnextchar [ {\mn@doi@}
  {\mn@doi@[]}}
\def\mn@doi@[#1]#2{\def\@tempa{#1}\ifx\@tempa\@empty \href
  {http://dx.doi.org/#2} {doi:#2}\else \href {http://dx.doi.org/#2} {#1}\fi
  \endgroup}
\def\mn@eprint#1#2{\mn@eprint@#1:#2::\@nil}
\def\mn@eprint@arXiv#1{\href {http://arxiv.org/abs/#1} {{\tt arXiv:#1}}}
\def\mn@eprint@dblp#1{\href {http://dblp.uni-trier.de/rec/bibtex/#1.xml}
  {dblp:#1}}
\def\mn@eprint@#1:#2:#3:#4\@nil{\def\@tempa {#1}\def\@tempb {#2}\def\@tempc
  {#3}\ifx \@tempc \@empty \let \@tempc \@tempb \let \@tempb \@tempa \fi \ifx
  \@tempb \@empty \def\@tempb {arXiv}\fi \@ifundefined
  {mn@eprint@\@tempb}{\@tempb:\@tempc}{\expandafter \expandafter \csname
  mn@eprint@\@tempb\endcsname \expandafter{\@tempc}}}

\bibitem[\protect\citeauthoryear{{Aarseth}}{{Aarseth}}{1973}]{aarseth_esc}
{Aarseth} S.~J.,  1973, \mn@doi [Vistas in Astronomy]
  {10.1016/0083-6656(73)90003-2}, \href
  {http://adsabs.harvard.edu/abs/1973VA.....15...13A} {15, 13}

\bibitem[\protect\citeauthoryear{{Aarseth}}{{Aarseth}}{2003}]{aarseth_book}
{Aarseth} S.~J.,  2003, {Gravitational N-Body Simulations}

\bibitem[\protect\citeauthoryear{{Aarseth}}{{Aarseth}}{2012}]{aarsethnb7}
{Aarseth} S.~J.,  2012, \mn@doi [\mnras] {10.1111/j.1365-2966.2012.20666.x},
  \href {http://adsabs.harvard.edu/abs/2012MNRAS.422..841A} {422, 841}

\bibitem[\protect\citeauthoryear{{Abbott} et~al.,}{{Abbott}
  et~al.}{2016a}]{abbott16a}
{Abbott} B.~P.,  et~al., 2016a, \mn@doi [Physical Review Letters]
  {10.1103/PhysRevLett.116.061102}, \href
  {http://adsabs.harvard.edu/abs/2016PhRvL.116f1102A} {116, 061102}

\bibitem[\protect\citeauthoryear{{Abbott} et~al.,}{{Abbott}
  et~al.}{2016b}]{abbott16b}
{Abbott} B.~P.,  et~al., 2016b, \mn@doi [Physical Review Letters]
  {10.1103/PhysRevLett.116.241103}, \href
  {http://adsabs.harvard.edu/abs/2016PhRvL.116x1103A} {116, 241103}

\bibitem[\protect\citeauthoryear{{Abbott} et~al.,}{{Abbott}
  et~al.}{2016c}]{abbrate16}
{Abbott} B.~P.,  et~al., 2016c, \mn@doi [\apjl] {10.3847/2041-8205/833/1/L1},
  \href {http://adsabs.harvard.edu/abs/2016ApJ...833L...1A} {833, L1}

\bibitem[\protect\citeauthoryear{{Abbott} et~al.,}{{Abbott}
  et~al.}{2017a}]{abbott17}
{Abbott} B.~P.,  et~al., 2017a, \mn@doi [Physical Review Letters]
  {10.1103/PhysRevLett.118.221101}, \href
  {http://adsabs.harvard.edu/abs/2017PhRvL.118v1101A} {118, 221101}

\bibitem[\protect\citeauthoryear{{Abbott} et~al.,}{{Abbott}
  et~al.}{2017b}]{abbott17b}
{Abbott} B.~P.,  et~al., 2017b, \mn@doi [Physical Review Letters]
  {10.1103/PhysRevLett.119.141101}, \href
  {http://adsabs.harvard.edu/abs/2017PhRvL.119n1101A} {119, 141101}

\bibitem[\protect\citeauthoryear{{Abbott} et~al.,}{{Abbott}
  et~al.}{2017c}]{abbott17c}
{Abbott} B.~P.,  et~al., 2017c, \mn@doi [Physical Review Letters]
  {10.1103/PhysRevLett.119.161101}, \href
  {http://adsabs.harvard.edu/abs/2017PhRvL.119p1101A} {119, 161101}

\bibitem[\protect\citeauthoryear{{Alexander}}{{Alexander}}{2017}]{tal17}
{Alexander} T.,  2017, \mn@doi [\araa] {10.1146/annurev-astro-091916-055306},
  \href {http://adsabs.harvard.edu/abs/2017ARA%26A..55...17A} {55, 17}

\bibitem[\protect\citeauthoryear{{Amaro-Seoane} \& {Chen}}{{Amaro-Seoane} \&
  {Chen}}{2016}]{Amaro-SeoaneChen2016}
{Amaro-Seoane} P.,  {Chen} X.,  2016, \mn@doi [\mnras] {10.1093/mnras/stw503},
  \href {http://adsabs.harvard.edu/abs/2016MNRAS.458.3075A} {458, 3075}

\bibitem[\protect\citeauthoryear{{Antonini} \& {Rasio}}{{Antonini} \&
  {Rasio}}{2016}]{antoninirasio2016}
{Antonini} F.,  {Rasio} F.~A.,  2016, \mn@doi [\apj]
  {10.3847/0004-637X/831/2/187}, \href
  {http://adsabs.harvard.edu/abs/2016ApJ...831..187A} {831, 187}

\bibitem[\protect\citeauthoryear{{Antonini}, {Chatterjee}, {Rodriguez},
  {Morscher}, {Pattabiraman}, {Kalogera}  \& {Rasio}}{{Antonini}
  et~al.}{2016}]{antonini16}
{Antonini} F.,  {Chatterjee} S.,  {Rodriguez} C.~L.,  {Morscher} M.,
  {Pattabiraman} B.,  {Kalogera} V.,   {Rasio} F.~A.,  2016, \mn@doi [\apj]
  {10.3847/0004-637X/816/2/65}, \href
  {http://adsabs.harvard.edu/abs/2016ApJ...816...65A} {816, 65}

\bibitem[\protect\citeauthoryear{{Arca-Sedda} \&
  {Capuzzo-Dolcetta}}{{Arca-Sedda} \& {Capuzzo-Dolcetta}}{2017a}]{megan17}
{Arca-Sedda} M.,  {Capuzzo-Dolcetta} R.,  2017a, preprint, \href
  {http://adsabs.harvard.edu/abs/2017arXiv170905567A} {} (\mn@eprint {arXiv}
  {1709.05567})

\bibitem[\protect\citeauthoryear{{Arca-Sedda} \&
  {Capuzzo-Dolcetta}}{{Arca-Sedda} \& {Capuzzo-Dolcetta}}{2017b}]{megan1}
{Arca-Sedda} M.,  {Capuzzo-Dolcetta} R.,  2017b, \mn@doi [\mnras]
  {10.1093/mnras/stx1586}, \href
  {http://adsabs.harvard.edu/abs/2017MNRAS.471..478A} {471, 478}

\bibitem[\protect\citeauthoryear{{Arca-Sedda}, {Capuzzo-Dolcetta}, {Antonini}
  \& {Seth}}{{Arca-Sedda} et~al.}{2015}]{ASCD15He}
{Arca-Sedda} M.,  {Capuzzo-Dolcetta} R.,  {Antonini} F.,   {Seth} A.,  2015,
  \mn@doi [\apj] {10.1088/0004-637X/806/2/220}, \href
  {http://adsabs.harvard.edu/abs/2015ApJ...806..220A} {806, 220}

\bibitem[\protect\citeauthoryear{{Arca-Sedda}, {Li}  \& {Kocsis}}{{Arca-Sedda}
  et~al.}{2018}]{ASKL18}
{Arca-Sedda} M.,  {Li} G.,   {Kocsis} B.,  2018, preprint, \href
  {http://adsabs.harvard.edu/abs/2018arXiv180506458A} {} (\mn@eprint {arXiv}
  {1805.06458})

\bibitem[\protect\citeauthoryear{{Askar}, {Szkudlarek}, {Gondek-Rosi{\'n}ska},
  {Giersz}  \& {Bulik}}{{Askar} et~al.}{2017}]{askar17}
{Askar} A.,  {Szkudlarek} M.,  {Gondek-Rosi{\'n}ska} D.,  {Giersz} M.,
  {Bulik} T.,  2017, \mn@doi [\mnras] {10.1093/mnrasl/slw177}, \href
  {http://adsabs.harvard.edu/abs/2017MNRAS.464L..36A} {464, L36}

\bibitem[\protect\citeauthoryear{{Bae}, {Kim}  \& {Lee}}{{Bae}
  et~al.}{2014}]{bae14}
{Bae} Y.-B.,  {Kim} C.,   {Lee} H.~M.,  2014, \mn@doi [\mnras]
  {10.1093/mnras/stu381}, \href
  {http://adsabs.harvard.edu/abs/2014MNRAS.440.2714B} {440, 2714}

\bibitem[\protect\citeauthoryear{{Baker}, {Centrella}, {Choi}, {Koppitz}, {van
  Meter}  \& {Miller}}{{Baker} et~al.}{2006}]{BakerEtAl2006}
{Baker} J.~G.,  {Centrella} J.,  {Choi} D.-I.,  {Koppitz} M.,  {van Meter}
  J.~R.,   {Miller} M.~C.,  2006, \mn@doi [ApJ] {10.1086/510448}, \href
  {http://adsabs.harvard.edu/abs/2006ApJ...653L..93B} {653, L93}

\bibitem[\protect\citeauthoryear{{Banerjee}}{{Banerjee}}{2017}]{sambaran1}
{Banerjee} S.,  2017, \mn@doi [\mnras] {10.1093/mnras/stw3392}, \href
  {http://adsabs.harvard.edu/abs/2017MNRAS.467..524B} {467, 524}

\bibitem[\protect\citeauthoryear{{Banerjee}}{{Banerjee}}{2018a}]{sambaran3}
{Banerjee} S.,  2018a, \mn@doi [\mnras] {10.1093/mnras/sty2608}, \href
  {http://adsabs.harvard.edu/abs/2018MNRAS.tmp.2494B} {}

\bibitem[\protect\citeauthoryear{{Banerjee}}{{Banerjee}}{2018b}]{sambaran2}
{Banerjee} S.,  2018b, \mn@doi [\mnras] {10.1093/mnras/stx2347}, \href
  {http://adsabs.harvard.edu/abs/2018MNRAS.473..909B} {473, 909}

\bibitem[\protect\citeauthoryear{{Banerjee}, {Baumgardt}  \&
  {Kroupa}}{{Banerjee} et~al.}{2010}]{sambaran10}
{Banerjee} S.,  {Baumgardt} H.,   {Kroupa} P.,  2010, \mn@doi [\mnras]
  {10.1111/j.1365-2966.2009.15880.x}, \href
  {http://adsabs.harvard.edu/abs/2010MNRAS.402..371B} {402, 371}

\bibitem[\protect\citeauthoryear{{Belczynski}, {Kalogera}  \&
  {Bulik}}{{Belczynski} et~al.}{2002}]{belczynski2002}
{Belczynski} K.,  {Kalogera} V.,   {Bulik} T.,  2002, \mn@doi [\apj]
  {10.1086/340304}, \href {http://adsabs.harvard.edu/abs/2002ApJ...572..407B}
  {572, 407}

\bibitem[\protect\citeauthoryear{{Belczynski}, {Bulik}, {Fryer}, {Ruiter},
  {Valsecchi}, {Vink}  \& {Hurley}}{{Belczynski} et~al.}{2010a}]{belczy2010}
{Belczynski} K.,  {Bulik} T.,  {Fryer} C.~L.,  {Ruiter} A.,  {Valsecchi} F.,
  {Vink} J.~S.,   {Hurley} J.~R.,  2010a, \mn@doi [\apj]
  {10.1088/0004-637X/714/2/1217}, \href
  {http://adsabs.harvard.edu/abs/2010ApJ...714.1217B} {714, 1217}

\bibitem[\protect\citeauthoryear{{Belczynski}, {Bulik}, {Fryer}, {Ruiter},
  {Valsecchi}, {Vink}  \& {Hurley}}{{Belczynski}
  et~al.}{2010b}]{belckzynski2010}
{Belczynski} K.,  {Bulik} T.,  {Fryer} C.~L.,  {Ruiter} A.,  {Valsecchi} F.,
  {Vink} J.~S.,   {Hurley} J.~R.,  2010b, \mn@doi [\apj]
  {10.1088/0004-637X/714/2/1217}, \href
  {http://adsabs.harvard.edu/abs/2010ApJ...714.1217B} {714, 1217}

\bibitem[\protect\citeauthoryear{{Belczynski}, {Bulik}, {Fryer}, {Ruiter},
  {Valsecchi}, {Vink}  \& {Hurley}}{{Belczynski} et~al.}{2010c}]{bel2010}
{Belczynski} K.,  {Bulik} T.,  {Fryer} C.~L.,  {Ruiter} A.,  {Valsecchi} F.,
  {Vink} J.~S.,   {Hurley} J.~R.,  2010c, \mn@doi [\apj]
  {10.1088/0004-637X/714/2/1217}, \href
  {http://cdsads.u-strasbg.fr/abs/2010ApJ...714.1217B} {714, 1217}

\bibitem[\protect\citeauthoryear{{Belczynski}, {Bulik}, {Fryer}, {Ruiter},
  {Valsecchi}, {Vink}  \& {Hurley}}{{Belczynski}
  et~al.}{2010d}]{belczynski2010}
{Belczynski} K.,  {Bulik} T.,  {Fryer} C.~L.,  {Ruiter} A.,  {Valsecchi} F.,
  {Vink} J.~S.,   {Hurley} J.~R.,  2010d, \mn@doi [ApJ]
  {10.1088/0004-637X/714/2/1217}, \href
  {http://adsabs.harvard.edu/abs/2010ApJ...714.1217B} {714, 1217}

\bibitem[\protect\citeauthoryear{{Benacquista} \& {Downing}}{{Benacquista} \&
  {Downing}}{2013}]{BenacquistaDowning2013}
{Benacquista} M.~J.,  {Downing} J.~M.~B.,  2013, \mn@doi [Living Reviews in
  Relativity] {10.12942/lrr-2013-4}, \href
  {http://adsabs.harvard.edu/abs/2013LRR....16....4B} {16, 4}

\bibitem[\protect\citeauthoryear{{Bethe} \& {Brown}}{{Bethe} \&
  {Brown}}{1998}]{bethe}
{Bethe} H.~A.,  {Brown} G.~E.,  1998, \mn@doi [\apj] {10.1086/306265}, \href
  {http://adsabs.harvard.edu/abs/1998ApJ...506..780B} {506, 780}

\bibitem[\protect\citeauthoryear{{Binney} \& {Tremaine}}{{Binney} \&
  {Tremaine}}{2008}]{bt}
{Binney} J.,  {Tremaine} S.,  2008, {Galactic Dynamics: Second Edition}.
Princeton University Press

\bibitem[\protect\citeauthoryear{{Campanelli}, {Lousto}, {Zlochower}  \&
  {Merritt}}{{Campanelli} et~al.}{2007}]{CampanelliEtAl2007}
{Campanelli} M.,  {Lousto} C.,  {Zlochower} Y.,   {Merritt} D.,  2007, \mn@doi
  [\apjl] {10.1086/516712}, \href
  {http://adsabs.harvard.edu/abs/2007ApJ...659L...5C} {659, L5}

\bibitem[\protect\citeauthoryear{{Chen} \& {Amaro-Seoane}}{{Chen} \&
  {Amaro-Seoane}}{2017}]{ChenAmaro-Seoane2017}
{Chen} X.,  {Amaro-Seoane} P.,  2017, \mn@doi [ApJ] {10.3847/2041-8213/aa74ce},
  \href {http://adsabs.harvard.edu/abs/2017ApJ...842L...2C} {842, L2}

\bibitem[\protect\citeauthoryear{{Dehnen}}{{Dehnen}}{1993}]{fbulge}
{Dehnen} W.,  1993, \mn@doi [\mnras] {10.1093/mnras/265.1.250}, \href
  {http://adsabs.harvard.edu/abs/1993MNRAS.265..250D} {265, 250}

\bibitem[\protect\citeauthoryear{{Downing}, {Benacquista}, {Giersz}  \&
  {Spurzem}}{{Downing} et~al.}{2010}]{downing2010}
{Downing} J.~M.~B.,  {Benacquista} M.~J.,  {Giersz} M.,   {Spurzem} R.,  2010,
  \mn@doi [\mnras] {10.1111/j.1365-2966.2010.17040.x}, \href
  {http://adsabs.harvard.edu/abs/2010MNRAS.407.1946D} {407, 1946}

\bibitem[\protect\citeauthoryear{{Duquennoy} \& {Mayor}}{{Duquennoy} \&
  {Mayor}}{1991}]{DuquennoyMayor1991}
{Duquennoy} A.,  {Mayor} M.,  1991, \aap, \href
  {http://adsabs.harvard.edu/abs/1991A%26A...248..485D} {248, 485}

\bibitem[\protect\citeauthoryear{{Fragione} \& {Kocsis}}{{Fragione} \&
  {Kocsis}}{2018}]{frak18}
{Fragione} G.,  {Kocsis} B.,  2018, preprint, \href
  {http://adsabs.harvard.edu/abs/2018arXiv180602351F} {} (\mn@eprint {arXiv}
  {1806.02351})

\bibitem[\protect\citeauthoryear{{Fragione} \& {Leigh}}{{Fragione} \&
  {Leigh}}{2018}]{fraglei18}
{Fragione} G.,  {Leigh} N.,  2018, \mn@doi [\mnras] {10.1093/mnras/sty1600},
  \href {http://adsabs.harvard.edu/abs/2018arXiv180310107F} {479, 3181}

\bibitem[\protect\citeauthoryear{{Fragione}, {Leigh}, {Ginsburg}  \&
  {Kocsis}}{{Fragione} et~al.}{2018a}]{frlgk18}
{Fragione} G.,  {Leigh} N.,  {Ginsburg} I.,   {Kocsis} B.,  2018a,
  arXiv:1806.08385, \href {http://adsabs.harvard.edu/abs/2018arXiv180608385F}
  {}

\bibitem[\protect\citeauthoryear{{Fragione}, {Ginsburg}  \&
  {Kocsis}}{{Fragione} et~al.}{2018b}]{fragk17}
{Fragione} G.,  {Ginsburg} I.,   {Kocsis} B.,  2018b, \mn@doi [\apj]
  {10.3847/1538-4357/aab368}, \href
  {http://adsabs.harvard.edu/abs/2017arXiv171100483F} {856, 92}

\bibitem[\protect\citeauthoryear{{Fregeau}, {Cheung}, {Portegies Zwart}  \&
  {Rasio}}{{Fregeau} et~al.}{2004}]{fregeau2004}
{Fregeau} J.~M.,  {Cheung} P.,  {Portegies Zwart} S.~F.,   {Rasio} F.~A.,
  2004, \mn@doi [\mnras] {10.1111/j.1365-2966.2004.07914.x}, \href
  {http://adsabs.harvard.edu/abs/2004MNRAS.352....1F} {352, 1}

\bibitem[\protect\citeauthoryear{{Fryer}, {Belczynski}, {Wiktorowicz},
  {Dominik}, {Kalogera}  \& {Holz}}{{Fryer} et~al.}{2012}]{fryer12}
{Fryer} C.~L.,  {Belczynski} K.,  {Wiktorowicz} G.,  {Dominik} M.,  {Kalogera}
  V.,   {Holz} D.~E.,  2012, \mn@doi [\apj] {10.1088/0004-637X/749/1/91}, \href
  {http://adsabs.harvard.edu/abs/2012ApJ...749...91F} {749, 91}

\bibitem[\protect\citeauthoryear{{Giacobbo} \& {Mapelli}}{{Giacobbo} \&
  {Mapelli}}{2018}]{giacobbo18}
{Giacobbo} N.,  {Mapelli} M.,  2018, \mn@doi [\mnras] {10.1093/mnras/sty1999},
  \href {http://adsabs.harvard.edu/abs/2018MNRAS.480.2011G} {480, 2011}

\bibitem[\protect\citeauthoryear{{Gonz{\'a}lez}, {Sperhake}, {Br{\"u}gmann},
  {Hannam}  \& {Husa}}{{Gonz{\'a}lez} et~al.}{2007}]{GonzalezEtAl2007}
{Gonz{\'a}lez} J.~A.,  {Sperhake} U.,  {Br{\"u}gmann} B.,  {Hannam} M.,
  {Husa} S.,  2007, \mn@doi [Physical Review Letters]
  {10.1103/PhysRevLett.98.091101}, \href
  {http://adsabs.harvard.edu/abs/2007PhRvL..98i1101G} {98, 091101}

\bibitem[\protect\citeauthoryear{{Goswami}, {Kiel}  \& {Rasio}}{{Goswami}
  et~al.}{2014}]{goswami2004}
{Goswami} S.,  {Kiel} P.,   {Rasio} F.~A.,  2014, \mn@doi [\apj]
  {10.1088/0004-637X/781/2/81}, \href
  {http://adsabs.harvard.edu/abs/2014ApJ...781...81G} {781, 81}

\bibitem[\protect\citeauthoryear{{Heger}, {Fryer}, {Woosley}, {Langer}  \&
  {Hartmann}}{{Heger} et~al.}{2003}]{heger2003}
{Heger} A.,  {Fryer} C.~L.,  {Woosley} S.~E.,  {Langer} N.,   {Hartmann} D.~H.,
   2003, \mn@doi [\apj] {10.1086/375341}, \href
  {http://adsabs.harvard.edu/abs/2003ApJ...591..288H} {591, 288}

\bibitem[\protect\citeauthoryear{{Heggie}}{{Heggie}}{1975}]{heggie75}
{Heggie} D.~C.,  1975, \mnras, \href
  {http://adsabs.harvard.edu/abs/1975MNRAS.173..729H} {173, 729}

\bibitem[\protect\citeauthoryear{{Hills}}{{Hills}}{1975}]{hills75}
{Hills} J.~G.,  1975, \mn@doi [\nat] {10.1038/254295a0}, \href
  {http://adsabs.harvard.edu/abs/1975Natur.254..295H} {254, 295}

\bibitem[\protect\citeauthoryear{{Hurley}, {Pols}  \& {Tout}}{{Hurley}
  et~al.}{2000}]{hurley2000}
{Hurley} J.~R.,  {Pols} O.~R.,   {Tout} C.~A.,  2000, \mn@doi [\mnras]
  {10.1046/j.1365-8711.2000.03426.x}, \href
  {http://adsabs.harvard.edu/abs/2000MNRAS.315..543H} {315, 543}

\bibitem[\protect\citeauthoryear{{Hurley}, {Tout}  \& {Pols}}{{Hurley}
  et~al.}{2002}]{hurley2002}
{Hurley} J.~R.,  {Tout} C.~A.,   {Pols} O.~R.,  2002, \mn@doi [\mnras]
  {10.1046/j.1365-8711.2002.05038.x}, \href
  {http://adsabs.harvard.edu/abs/2002MNRAS.329..897H} {329, 897}

\bibitem[\protect\citeauthoryear{{Janka}}{{Janka}}{2013}]{Janka2013}
{Janka} H.-T.,  2013, \mn@doi [\mnras] {10.1093/mnras/stt1106}, \href
  {http://adsabs.harvard.edu/abs/2013MNRAS.434.1355J} {434, 1355}

\bibitem[\protect\citeauthoryear{{Kimpson}, {Spera}, {Mapelli}  \&
  {Ziosi}}{{Kimpson} et~al.}{2016}]{kimpson}
{Kimpson} T.~O.,  {Spera} M.,  {Mapelli} M.,   {Ziosi} B.~M.,  2016, \mn@doi
  [\mnras] {10.1093/mnras/stw2085}, \href
  {http://adsabs.harvard.edu/abs/2016MNRAS.463.2443K} {463, 2443}

\bibitem[\protect\citeauthoryear{{Kozai}}{{Kozai}}{1962}]{kozai}
{Kozai} Y.,  1962, \mn@doi [\aj] {10.1086/108790}, \href
  {http://cdsads.u-strasbg.fr/abs/1962AJ.....67..591K} {67, 591}

\bibitem[\protect\citeauthoryear{{Kroupa}}{{Kroupa}}{2001}]{kroupa01}
{Kroupa} P.,  2001, \mn@doi [\mnras] {10.1046/j.1365-8711.2001.04022.x}, \href
  {http://adsabs.harvard.edu/abs/2001MNRAS.322..231K} {322, 231}

\bibitem[\protect\citeauthoryear{{Kupi}, {Amaro-Seoane}  \& {Spurzem}}{{Kupi}
  et~al.}{2006}]{KupiEtAl06}
{Kupi} G.,  {Amaro-Seoane} P.,   {Spurzem} R.,  2006, \mn@doi []
  {10.1111/j.1745-3933.2006.00205.x}, \href
  {http://adsabs.harvard.edu/cgi-bin/nph-bib_query?bibcode=2006MNRAS.tmpL..77K&db_key=AST}
  {pp~L77+}

\bibitem[\protect\citeauthoryear{{Lidov}}{{Lidov}}{1962}]{lidov}
{Lidov} M.~L.,  1962, \mn@doi [\planss] {10.1016/0032-0633(62)90129-0}, \href
  {http://cdsads.u-strasbg.fr/abs/1962P%26SS....9..719L} {9, 719}

\bibitem[\protect\citeauthoryear{{Loeb}}{{Loeb}}{2016}]{loeb16}
{Loeb} A.,  2016, \mn@doi [\apjl] {10.3847/2041-8205/819/2/L21}, \href
  {http://adsabs.harvard.edu/abs/2016ApJ...819L..21L} {819, L21}

\bibitem[\protect\citeauthoryear{{Mapelli}}{{Mapelli}}{2016}]{mapelli16}
{Mapelli} M.,  2016, \mn@doi [\mnras] {10.1093/mnras/stw869}, \href
  {http://adsabs.harvard.edu/abs/2016MNRAS.459.3432M} {459, 3432}

\bibitem[\protect\citeauthoryear{{Mapelli} \& {Bressan}}{{Mapelli} \&
  {Bressan}}{2013}]{mapelli13}
{Mapelli} M.,  {Bressan} A.,  2013, \mn@doi [\mnras] {10.1093/mnras/stt119},
  \href {http://adsabs.harvard.edu/abs/2013MNRAS.430.3120M} {430, 3120}

\bibitem[\protect\citeauthoryear{{Mapelli}, {Moore}, {Giordano}, {Mayer},
  {Colpi}, {Ripamonti}  \& {Callegari}}{{Mapelli} et~al.}{2008}]{mapelli08}
{Mapelli} M.,  {Moore} B.,  {Giordano} L.,  {Mayer} L.,  {Colpi} M.,
  {Ripamonti} E.,   {Callegari} S.,  2008, \mn@doi [\mnras]
  {10.1111/j.1365-2966.2007.12534.x}, \href
  {http://adsabs.harvard.edu/abs/2008MNRAS.383..230M} {383, 230}

\bibitem[\protect\citeauthoryear{{Mapelli}, {Huwyler}, {Mayer}, {Jetzer}  \&
  {Vecchio}}{{Mapelli} et~al.}{2010}]{mapelli10}
{Mapelli} M.,  {Huwyler} C.,  {Mayer} L.,  {Jetzer} P.,   {Vecchio} A.,  2010,
  \mn@doi [\apj] {10.1088/0004-637X/719/2/987}, \href
  {http://adsabs.harvard.edu/abs/2010ApJ...719..987M} {719, 987}

\bibitem[\protect\citeauthoryear{{Mathieu}}{{Mathieu}}{2008}]{Mathieu2008}
{Mathieu} R.~D.,  2008, in {Vesperini} E.,  {Giersz} M.,   {Sills} A.,  eds,
  IAU Symposium Vol. 246, Dynamical Evolution of Dense Stellar Systems. pp
  79--88, \mn@doi{10.1017/S1743921308015366}

\bibitem[\protect\citeauthoryear{{Mikkola} \& {Merritt}}{{Mikkola} \&
  {Merritt}}{2008}]{mikkola08}
{Mikkola} S.,  {Merritt} D.,  2008, \mn@doi [\aj]
  {10.1088/0004-6256/135/6/2398}, \href
  {http://adsabs.harvard.edu/abs/2008AJ....135.2398M} {135, 2398}

\bibitem[\protect\citeauthoryear{{Mikkola} \& {Tanikawa}}{{Mikkola} \&
  {Tanikawa}}{1999}]{mikkola99}
{Mikkola} S.,  {Tanikawa} K.,  1999, \mn@doi [\mnras]
  {10.1046/j.1365-8711.1999.02982.x}, \href
  {http://adsabs.harvard.edu/abs/1999MNRAS.310..745M} {310, 745}

\bibitem[\protect\citeauthoryear{{Miyamoto} \& {Nagai}}{{Miyamoto} \&
  {Nagai}}{1975}]{fdisk}
{Miyamoto} M.,  {Nagai} R.,  1975, \pasj, \href
  {http://adsabs.harvard.edu/abs/1975PASJ...27..533M} {27, 533}

\bibitem[\protect\citeauthoryear{{Naoz}}{{Naoz}}{2016}]{naoz}
{Naoz} S.,  2016, \mn@doi [\araa] {10.1146/annurev-astro-081915-023315}, \href
  {http://cdsads.u-strasbg.fr/abs/2016ARA%26A..54..441N} {54, 441}

\bibitem[\protect\citeauthoryear{{Park}, {Kim}, {Lee}, {Bae}  \&
  {Belczynski}}{{Park} et~al.}{2017}]{park17}
{Park} D.,  {Kim} C.,  {Lee} H.~M.,  {Bae} Y.-B.,   {Belczynski} K.,  2017,
  \mn@doi [\mnras] {10.1093/mnras/stx1015}, \href
  {http://adsabs.harvard.edu/abs/2017MNRAS.469.4665P} {469, 4665}

\bibitem[\protect\citeauthoryear{{Perets}, {Li}, {Lombardi}  \&
  {Milcarek}}{{Perets} et~al.}{2016}]{perets16}
{Perets} H.~B.,  {Li} Z.,  {Lombardi} Jr. J.~C.,   {Milcarek} Jr. S.~R.,  2016,
  \mn@doi [\apj] {10.3847/0004-637X/823/2/113}, \href
  {http://adsabs.harvard.edu/abs/2016ApJ...823..113P} {823, 113}

\bibitem[\protect\citeauthoryear{{Perna}, {Chruslinska}, {Corsi}  \&
  {Belczynski}}{{Perna} et~al.}{2018}]{PernaEtAl2018}
{Perna} R.,  {Chruslinska} M.,  {Corsi} A.,   {Belczynski} K.,  2018, \mn@doi
  [\mnras] {10.1093/mnras/sty814}, \href
  {http://adsabs.harvard.edu/abs/2018MNRAS.477.4228P} {477, 4228}

\bibitem[\protect\citeauthoryear{Peters}{Peters}{1964}]{peters64}
Peters P.~C.,  1964, \mn@doi [Phys. Rev.] {10.1103/PhysRev.136.B1224}, 136,
  B1224

\bibitem[\protect\citeauthoryear{{Peters} \& {Mathews}}{{Peters} \&
  {Mathews}}{1963}]{PM63}
{Peters} P.~C.,  {Mathews} J.,  1963, Physical Review, 131, 435

\bibitem[\protect\citeauthoryear{{Piskunov}, {Schilbach}, {Kharchenko},
  {R{\"o}ser}  \& {Scholz}}{{Piskunov} et~al.}{2008}]{piskunov08}
{Piskunov} A.~E.,  {Schilbach} E.,  {Kharchenko} N.~V.,  {R{\"o}ser} S.,
  {Scholz} R.-D.,  2008, \mn@doi [\aap] {10.1051/0004-6361:20078525}, \href
  {http://adsabs.harvard.edu/abs/2008A%26A...477..165P} {477, 165}

\bibitem[\protect\citeauthoryear{{Plummer}}{{Plummer}}{1911}]{Plum}
{Plummer} H.~C.,  1911, MNRAS, \href
  {http://adsabs.harvard.edu/abs/1911MNRAS..71..460P} {71, 460}

\bibitem[\protect\citeauthoryear{{Podsiadlowski}, {Langer}, {Poelarends},
  {Rappaport}, {Heger}  \& {Pfahl}}{{Podsiadlowski} et~al.}{2004}]{podsi04}
{Podsiadlowski} P.,  {Langer} N.,  {Poelarends} A.~J.~T.,  {Rappaport} S.,
  {Heger} A.,   {Pfahl} E.,  2004, \mn@doi [\apj] {10.1086/421713}, \href
  {http://adsabs.harvard.edu/abs/2004ApJ...612.1044P} {612, 1044}

\bibitem[\protect\citeauthoryear{{Podsiadlowski}, {Pfahl}  \&
  {Rappaport}}{{Podsiadlowski} et~al.}{2005}]{podslia2005}
{Podsiadlowski} P.,  {Pfahl} E.,   {Rappaport} S.,  2005, in {Rasio} F.~A.,
  {Stairs} I.~H.,  eds,  Astronomical Society of the Pacific Conference Series
  Vol. 328, Binary Radio Pulsars. p.~327

\bibitem[\protect\citeauthoryear{{Portegies Zwart}, {McMillan}  \&
  {Gieles}}{{Portegies Zwart} et~al.}{2010}]{spz2010}
{Portegies Zwart} S.~F.,  {McMillan} S.~L.~W.,   {Gieles} M.,  2010, \mn@doi
  [\araa] {10.1146/annurev-astro-081309-130834}, \href
  {http://adsabs.harvard.edu/abs/2010ARA%26A..48..431P} {48, 431}

\bibitem[\protect\citeauthoryear{{Postnov} \& {Kuranov}}{{Postnov} \&
  {Kuranov}}{2017}]{postnov17}
{Postnov} K.,  {Kuranov} A.,  2017, preprint, \href
  {http://adsabs.harvard.edu/abs/2017arXiv170600369P} {} (\mn@eprint {arXiv}
  {1706.00369})

\bibitem[\protect\citeauthoryear{{Postnov} \& {Yungelson}}{{Postnov} \&
  {Yungelson}}{2014}]{postnov14}
{Postnov} K.~A.,  {Yungelson} L.~R.,  2014, \mn@doi [Living Reviews in
  Relativity] {10.12942/lrr-2014-3}, \href
  {http://adsabs.harvard.edu/abs/2014LRR....17....3P} {17, 3}

\bibitem[\protect\citeauthoryear{{Repetto}, {Davies}  \&
  {Sigurdsson}}{{Repetto} et~al.}{2012}]{RepettoEt2012}
{Repetto} S.,  {Davies} M.~B.,   {Sigurdsson} S.,  2012, \mn@doi [\mnras]
  {10.1111/j.1365-2966.2012.21549.x}, \href
  {http://adsabs.harvard.edu/abs/2012MNRAS.425.2799R} {425, 2799}

\bibitem[\protect\citeauthoryear{{Rodriguez}, {Morscher}, {Pattabiraman},
  {Chatterjee}, {Haster}  \& {Rasio}}{{Rodriguez} et~al.}{2015}]{rodriguez15}
{Rodriguez} C.~L.,  {Morscher} M.,  {Pattabiraman} B.,  {Chatterjee} S.,
  {Haster} C.-J.,   {Rasio} F.~A.,  2015, \mn@doi [Physical Review Letters]
  {10.1103/PhysRevLett.115.051101}, \href
  {http://adsabs.harvard.edu/abs/2015PhRvL.115e1101R} {115, 051101}

\bibitem[\protect\citeauthoryear{{Rodriguez}, {Chatterjee}  \&
  {Rasio}}{{Rodriguez} et~al.}{2016a}]{rodriguez16}
{Rodriguez} C.~L.,  {Chatterjee} S.,   {Rasio} F.~A.,  2016a, \mn@doi [\prd]
  {10.1103/PhysRevD.93.084029}, \href
  {http://adsabs.harvard.edu/abs/2016PhRvD..93h4029R} {93, 084029}

\bibitem[\protect\citeauthoryear{{Rodriguez}, {Haster}, {Chatterjee},
  {Kalogera}  \& {Rasio}}{{Rodriguez} et~al.}{2016b}]{rodriguez16b}
{Rodriguez} C.~L.,  {Haster} C.-J.,  {Chatterjee} S.,  {Kalogera} V.,   {Rasio}
  F.~A.,  2016b, \mn@doi [\apjl] {10.3847/2041-8205/824/1/L8}, \href
  {http://adsabs.harvard.edu/abs/2016ApJ...824L...8R} {824, L8}

\bibitem[\protect\citeauthoryear{{Rodriguez}, {Amaro-Seoane}, {Chatterjee}  \&
  {Rasio}}{{Rodriguez} et~al.}{2018}]{RodriguezEtAl2018}
{Rodriguez} C.~L.,  {Amaro-Seoane} P.,  {Chatterjee} S.,   {Rasio} F.~A.,
  2018, \mn@doi [Physical Review Letters] {10.1103/PhysRevLett.120.151101},
  \href {http://adsabs.harvard.edu/abs/2018PhRvL.120o1101R} {120, 151101}

\bibitem[\protect\citeauthoryear{{Spera} \& {Mapelli}}{{Spera} \&
  {Mapelli}}{2017}]{spema17}
{Spera} M.,  {Mapelli} M.,  2017, preprint, \href
  {http://adsabs.harvard.edu/abs/2017arXiv170606109S} {} (\mn@eprint {arXiv}
  {1706.06109})

\bibitem[\protect\citeauthoryear{{Spera}, {Mapelli}  \& {Bressan}}{{Spera}
  et~al.}{2015}]{speetal15}
{Spera} M.,  {Mapelli} M.,   {Bressan} A.,  2015, \mn@doi [\mnras]
  {10.1093/mnras/stv1161}, \href
  {http://adsabs.harvard.edu/abs/2015MNRAS.451.4086S} {451, 4086}

\bibitem[\protect\citeauthoryear{{Spitzer}}{{Spitzer}}{1987}]{spi87}
{Spitzer} L.,  1987, {Dynamical evolution of globular clusters}

\bibitem[\protect\citeauthoryear{{Stone} \& {Metzger}}{{Stone} \&
  {Metzger}}{2016}]{stonemetzger}
{Stone} N.~C.,  {Metzger} B.~D.,  2016, \mn@doi [\mnras]
  {10.1093/mnras/stv2281}, \href
  {http://adsabs.harvard.edu/abs/2016MNRAS.455..859S} {455, 859}

\bibitem[\protect\citeauthoryear{{Stone} \& {van Velzen}}{{Stone} \& {van
  Velzen}}{2016}]{stone16b}
{Stone} N.~C.,  {van Velzen} S.,  2016, \mn@doi [\apjl]
  {10.3847/2041-8205/825/1/L14}, \href
  {http://adsabs.harvard.edu/abs/2016ApJ...825L..14S} {825, L14}

\bibitem[\protect\citeauthoryear{{Stone}, {K{\"u}pper}  \& {Ostriker}}{{Stone}
  et~al.}{2017}]{stone17}
{Stone} N.~C.,  {K{\"u}pper} A.~H.~W.,   {Ostriker} J.~P.,  2017, \mn@doi
  [\mnras] {10.1093/mnras/stx097}, \href
  {http://adsabs.harvard.edu/abs/2017MNRAS.467.4180S} {467, 4180}

\bibitem[\protect\citeauthoryear{{The LIGO Scientific Collaboration}
  et~al.,}{{The LIGO Scientific Collaboration} et~al.}{2017}]{abbott17a}
{The LIGO Scientific Collaboration} et~al., 2017, preprint, \href
  {http://adsabs.harvard.edu/abs/2017arXiv171105578T} {} (\mn@eprint {arXiv}
  {1711.05578})

\bibitem[\protect\citeauthoryear{{Tutukov} \& {Cherepashchuk}}{{Tutukov} \&
  {Cherepashchuk}}{2017}]{tutukov17}
{Tutukov} A.~V.,  {Cherepashchuk} A.~M.,  2017, \mn@doi [Astronomy Reports]
  {10.1134/S1063772917100092}, \href
  {http://adsabs.harvard.edu/abs/2017ARep...61..833T} {61, 833}

\bibitem[\protect\citeauthoryear{{Tutukov} \& {Yungelson}}{{Tutukov} \&
  {Yungelson}}{1973}]{tutukov}
{Tutukov} A.,  {Yungelson} L.,  1973, Nauchnye Informatsii, \href
  {http://adsabs.harvard.edu/abs/1973NInfo..27...70T} {27, 70}

\bibitem[\protect\citeauthoryear{{Zampieri} \& {Roberts}}{{Zampieri} \&
  {Roberts}}{2009}]{zampieri09}
{Zampieri} L.,  {Roberts} T.~P.,  2009, \mn@doi [\mnras]
  {10.1111/j.1365-2966.2009.15509.x}, \href
  {http://adsabs.harvard.edu/abs/2009MNRAS.400..677Z} {400, 677}

\bibitem[\protect\citeauthoryear{{Ziosi}, {Mapelli}, {Branchesi}  \&
  {Tormen}}{{Ziosi} et~al.}{2014}]{ziosi2014}
{Ziosi} B.~M.,  {Mapelli} M.,  {Branchesi} M.,   {Tormen} G.,  2014, \mn@doi
  [\mnras] {10.1093/mnras/stu824}, \href
  {http://adsabs.harvard.edu/abs/2014MNRAS.441.3703Z} {441, 3703}

\makeatother
\end{thebibliography}

\label{lastpage}
\end{document}